\documentclass[a4paper,11pt]{article}

\pdfoutput=1 % if your are submitting a pdflatex (i.e. if you have
\usepackage{fullpage}

\usepackage{graphicx}

\usepackage{mathrsfs}
\usepackage{comment}
\usepackage{centernot}
\usepackage{mathtools}
\usepackage{stmaryrd}

\usepackage{latexsym,amssymb,amstext,amsmath}
\usepackage{hyperref}
\usepackage{graphbox,graphicx}
\usepackage{subcaption}
\usepackage{amsthm}
\usepackage{esint}
\usepackage{floatrow}
\usepackage{caption}
\usepackage{placeins}

\def\bea {\begin{eqnarray}}
\def\eea {\end{eqnarray}}

\def\be {\begin{equation}}
\def\ee {\end{equation}}

\def\beq{\begin{equation}}
\def\eeq{\end{equation}}
\def\beqa{\begin{eqnarray}}
\def\eeqa{\end{eqnarray}}

\newtheorem{theorem}{Theorem}[section]

\newtheorem{proposition}[theorem]{Proposition}
\newtheorem{corollary}[theorem]{Corollary}

\theoremstyle{definition}

\newtheorem{remark}[theorem]{Remark}

\newtheorem{example}[theorem]{Example}

\DeclareMathOperator{\Tr}{Tr}

\begin{document}

\title{\textbf{\LARGE Solving gravitational field equations by Wiener-Hopf matrix factorisation, and beyond}}

\author{M.~Cristina C\^amara and Gabriel Lopes Cardoso}
\date{\small 
\vspace{-5ex}
\begin{quote}
\emph{
\begin{itemize}
\item[]
Center for Mathematical Analysis, Geometry and Dynamical Systems,\\
  Department of Mathematics, 
  Instituto Superior T\'ecnico, Universidade de Lisboa,\\
  Av. Rovisco Pais, 1049-001 Lisboa, Portugal
  \end{itemize}
}
\end{quote}
{\tt 
cristina.camara@tecnico.ulisboa.pt, gabriel.lopes.cardoso@tecnico.ulisboa.pt}
}
\maketitle

\begin{abstract}
\noindent
By viewing Einstein’s field equations -- reduced to two dimensions -- as an integrable system, one can simultaneously obtain exact solutions to both the equations themselves and their associated Lax pair via a canonical Wiener–Hopf factorisation of a so-called monodromy matrix.
In this article, we review this remarkable interplay between gravitational field equations, integrable systems, Riemann–Hilbert problems, and Wiener–Hopf factorisation theory, with particular emphasis on developments from the past decade enabled by advances in Wiener–Hopf factorisation techniques
arising from the study of singular integral equations and Toeplitz operators. Through a variety of concrete examples, we illustrate how Wiener–Hopf factorisation yields explicit, exact solutions to the field equations of gravitational theories, and how its generalisation
through a so-called $\tau$-invariance property provides a new solution-generating method. Along the way, we aim to demonstrate the importance of an interdisciplinary approach -- grounded in General Relativity, Complex Analysis, and Operator Theory -- for the study of gravitational field equations.

\end{abstract}

%%%%%%
\section{Introduction \label{sec:intro}} 
%%%%%%

The study of the integrability of Einstein’s field equations, and their generalisations to the field 
equations of other gravitational theories, has evolved into a subject with a long and rich history 
(see, for instance, \cite{Klein:2005rne} and the recent review articles \cite{Woodhouse1997,Alekseev:2010mx,Korotkin:2023lrg}). As 
emphasized in these reviews, a variety of methods have been developed to search for exact
solutions of Einstein’s field equations, both in vacuum and in the presence of a Maxwell field.
In the late 1970's and early 1980's, several solution generating techniques for the four-dimensional 
vacuum Einstein field equations and Einstein-Maxwell equations were explored - for a survey of exact solutions, see
\cite{Griffiths:2009dfa,Stephani:2003tm}.

Many of these methods are based on Geroch’s observation, in his study of the two-
dimensional 
equations 
governing stationary axisymmetric solutions of Einstein’s vacuum field equations \cite{Geroch:1970nt,Geroch:1972yt}, that each solution is 
associated with an infinite family of potentials. From these, Geroch was able to generate, starting from 
a given "seed” solution, a family of new solutions to the equations. He conjectured that any stationary, 
axisymmetric solution could be obtained from Minkowski space-time via an infinite-dimensional group 
of transformations - a conjecture later proven by Ernst and Hauser \cite{10.1063/1.525012}. However, a concrete procedure
for carrying out these transformations on a given "seed" metric was not provided.

In a series of papers \cite{Kinnersley:1977pg,Kinnersley:1977ph,Kinnersley:1978pz,10.1063/1.523580}, Kinnersley and Chitre succeeded in performing these transformations in 
certain
cases, using a $2 \times 2$ matrix generating function depending on Weyl coordinates and a complex 
parameter, 
which was required to satisfy a system of partial differential equations (PDE's). 

In 1978, in a short paper \cite{PhysRevLett.41.521}, Maison conjectured that Einstein's field equations in the
stationary axially symmetric case constituted a "completely integrable Hamiltonian system", based on the
existence of a linear eigenvalue problem "in the spirit of Lax", with the non-linear system of Einstein's 
field equations appearing as its compatibility condition.

Shortly after, Belinski and Zakharov, taking a completely different approach, proved the 
integrability of the vacuum Einstein field equations in the presence of two commuting isometries \cite{Belinsky:1971nt,Belinsky:1979mh}. They
constructed an overdetermined system of linear partial differential equations depending on a complex 
spectral
parameter 
$\tau$,  as well as on two of the space-time coordinates, and demonstrated that Einstein’s field equations in vacuum
arise as the compatibility condition of this linear system.

One of the most successful techniques for generating exact solutions to Einstein's field equations, pioneered by Belinski and Zakharov, is 
the 
inverse scattering method \cite{abseg}.
In their seminal 
papers \cite{Belinsky:1971nt,Belinsky:1979mh}, they developed a practical approach based on a modified version of the inverse scattering 
problem \cite{Zakharov1970ExactTO,zakharov1980integrability}. 
This method enables the explicit construction of various classes of new exact solutions to 
Einstein’s field equations in vacuum
in cases where the metric tensor depends on only two variables, provided that a "seed"
solution is known. The approach involves solving an associated linear system, assuming 
the
existence of a known solution $\psi_0$. A new solution $\psi$ is then constructed via the transformation
$\psi = \chi \, \psi_0$, where $\chi$ is a matrix of a specific form that must be determined such that 
$\psi$
also satisfies the linear system. This requirement leads to a new linear system of differential equations 
for
$\chi$, the solution of which reduces to solving a Riemann-Hilbert problem on a certain circle 
in 
the complex plane \cite{Belinsky:1971nt}. 
Once $\psi$ is determined, the corresponding solution to the original nonlinear Einstein's field equations 
can
be extracted by evaluating $\psi$ at $\tau = 0$, where $\tau$ is the
complex spectral parameter. 

One difficulty with the inverse scattering method, however, is that it does not necessarily reproduce all the desired features of the solutions -- particularly the group structure in models where a symmetry group 
$G$  acts as a solution generating group \cite{Figueras:2009mc}.

This issue does not arise in the alternative group-theoretical framework proposed by Breitenlohner and 
Maison in \cite{Breitenlohner:1986um}. Their aim was to provide a clear group-theoretical understanding of various solution 
generating
techniques and to demonstrate that the action of the Geroch group is directly related to the inverse
scattering method developed for completely integrable systems. Focusing on the simplest case, pure 
Einstein
gravity in four dimensions, they proposed a method for obtaining solutions via a so-called linear spectral problem, 
in which
Einstein's field equations in vacuum emerge as compatibility conditions. This approach also involves 
solving a
specific Riemann-Hilbert problem and acting with the infinite dimensional Geroch group on a solution of the linear spectral problem.
However, this method is not easy to implement 
\cite{Breitenlohner:1986um,Figueras:2009mc,Katsimpouri:2012ky,Katsimpouri:2013wka,Chakrabarty:2014ora,Katsimpouri:2014ara}.

A completely different method to obtaining stationary axisymmetric
solutions to Einstein's field equations in vacuum, different from 
the ``B\"acklund transformation" approach often adopted to obtain new solutions to integrable PDE's  \cite{abseg,Aktosun}, is due to 
Ward in \cite{Ward1982}. Motivated by Penrose's construction for vacuum spaces with 
self-dual curvature tensor \cite{Penrose:1976js}, 
Ward's approach
translates the problem into one of complex geometry, using
twistor theory and a correspondence between solutions of the field equations and complex vector bundles.

One way to describe vector bundles is to specify
a "patching matrix", whose elements depend on a
complex variable $\tau$ as well as on the space-time coordinates $(\rho,v)$. 
Ward's construction consists in "splitting", i.e.
factorising a $2 \times 2$ "patching matrix" of the form
 \bea
H(\tau, \rho, v) = \begin{pmatrix}
    h_1 (\omega) & (- \tau)^k \, h_2 (\omega) \\
    \tau^{-k}  \, h_2 (\omega) & h_3 (\omega)
\end{pmatrix} \quad \text{with} \quad \omega= v + \frac{\rho}{2}  \left( \frac{ 1 - \tau^2}{\tau} \right) \;\;\;,\;\;\; k \in \mathbb{Z} \;,
\eea
as
\bea
H(\tau, \rho, v) = {\hat H} (\tau, \rho, v) \, H^{-1} (\tau, \rho,v) \;,
\label{splitF}
\eea
with ${\hat H}, H$ non singular, and $H$ analytic with respect to $\tau$ for $|\tau| \leq 1$,
${\hat H}$ analytic with respect to $\tau$ for $|\tau| \geq 1$, including the point $\infty$.

The solution $J(\rho,v)$ of Einstein's field equations in vacuum, written in signature $(+,-,-,-)$ in the form
\bea
ds^2_4 = \rho J_{ij} dy^i dy^j - \Omega (d\rho^2 + dv^2) \;,
\eea
where $y^1, y^2, \rho, v$ are the space-time coordinates, $\Omega = \Omega (\rho,v) > 0$ is a function determined by integration once $J$ is known, and $J$ is a symmetric $2 \times 2$
matrix of real valued functions of $\rho$ and $v$ with $\det J = -1$, 
is then obtained from \eqref{splitF} by the equality
\bea
J = P \, H(0, \rho, v) \,  {\hat H}^{-1} (\infty, \rho,v) \, P \;,
\eea
where $P = {\rm diag} \left( \rho^{-k/2}, \rho^{k/2} \right)$ (see Appendix \ref{sec:wardcan}). 

The main difficulty in this construction procedure,
as observed by Ward in \cite{Ward1982}, is that of finding the matrices ${\hat H}$ and $H$ in 
the factorisation \eqref{splitF}, since there 
is no systematic procedure for obtaining it. Nevertheless, the twistor approach 
to stationary axisymmetric space-times 
was illustrated by Ward in several examples, including the Schwarzschild solution and the family of 
Harrison's metrics \cite{Harrison}.

Ward's paper was the first to recast the stationary axisymmetric vacuum case of Einsteins's field
equations as a Riemann-Hilbert factorisation problem,  without resorting to a
previously known "seed solution" nor to B\"acklund transformations, and showing that vacuum solutions
of Einstein's field equations  can be viewed as vector bundles over
related twistor spaces in such a way that the
space-time metric can be recovered from the "patching matrix" by solving a Riemann-Hilbert factorisation 
problem.

Ward's twistor analysis of stationary axisymmetric solutions motivated much subsequent work
following the ideas introduced in his seminal paper \cite{Ward1982}. 
This approach was further developed in \cite{Woodhouse1988}, where the connection to solution-generating 
techniques in General Relativity is explored, and twistor theory is used to explain the appearance of 
Riemann-Hilbert problems in the construction of exact solutions. In addition, Ward’s framework was 
applied to gravitational waves with cylindrical symmetry in \cite{woodh1989} (see 
\cite{Fletcher_Woodhouse_1990} for a review of solutions generated by Ward's technique).

As mentioned above, linear systems and Riemann-Hilbert problems lie at the core of several methods for generating solutions to gravitational field equations. 
The main difficulty in applying these methods, as emphasized by many authors, see for example  \cite{Ward1982, Breitenlohner:1986um, Katsimpouri:2012ky}, lies in solving the associated Riemann-Hilbert 
problems -- particularly those involving factorisations of the type \eqref{splitF} -- in a sufficiently explicit form as to allow clear information about the solution to be extracted. However, 
recent developments in Riemann-Hilbert methods and advances in explicit Wiener-Hopf factorisation techniques (including numerical methods and RH problems on Riemann surfaces), in connection with the study of Toeplitz operators \cite{ccarp,BS,RogMish,KAMR,adu,CSSlax,CSSrs,CCsym,CDR,ES,mish}, have allowed to overcome many difficulties
and to obtain new results and solutions in the past decade. These have not, to the authors' knowledge, been covered in comprehensive surveys. In this paper we  therefore review how Riemann–Hilbert problems, in conjunction with Wiener-Hopf matrix factorisation 
methods grounded in complex analysis and operator theory, can be employed to construct {\it explicit} solutions to both the gravitational field equations, seen as an integrable system, and the 
underlying linear system.
The linear system under consideration is the one introduced in \cite{Lu:2007jc}, which is equivalent to the one formulated by Breitenlohner and Maison in \cite{Breitenlohner:1986um}. We will encounter two distinct
Riemann-Hilbert problems: one that we refer to as the {\it Riemann–Hilbert factorisation problem}, and another that we call the {\it  injectivity (vectorial) Riemann–Hilbert problem}. 
The latter is obtained by applying results from operator theory and is crucial in answering the question of {\it existence} of a canonical Wiener-Hopf factorisation.

This review is intended for researchers from two distinct communities, namely General Relativity on the one hand, and Complex Analysis  and Operator Theory, with an interest in applications in Mathematical Physics, on the other. 
In addition, by providing explicit examples,
we aim to offer a text that can serve as an introduction for graduate students and researchers to the topics discussed in the review and to their interrelations.
Its structure is as follows. In Section  \ref{lsc}, we begin by reviewing the formulation of the gravitational field equations as an integrable system. We introduce the Breitenlohner-Maison linear system, whose formulation 
involves an algebraic curve called the spectral curve. In Section \ref{sec:RHP} we review Riemann-Hilbert problems, Wiener-Hopf factorisation techniques and 
define admissible contours. We then discuss how canonical Wiener-Hopf factorisations of monodromy matrices with respect to admissible contours
provide solutions to both the Breitenlohner-Maison linear system as well as to its compatibility equations, the gravitational field equations.
We illustrate how the canonical factorisation of the same monodromy matrix, with respect to different admissible contours, can give rise to distinct solutions of the gravitational field equations.
In Section \ref{sec:bwh} we turn to a discussion of the existence of a canonical Wiener-Hopf factorisation. This is addressed via the connection with Toeplitz operators and results from
Operator Theory. As an example, we 
discuss the breakdown of the existence of a canonical Wiener-Hopf factorisation when approaching the ergosurface of the Kerr black hole of General Relatvity. In Section \ref{sec:tauinv} we discuss a recently introduced
solution generation method by multiplication, based on an 
approach called $\tau$-invariance, which enables us to 
go beyond canonical Wiener-Hopf factorisation in constructing solutions of the gravitational field equations.
We illustrate this method with several examples. Section \ref{sec:exam} presents several additional illustrative examples. In Section \ref{sec:open} we formulate a list of open questions related to the topics discussed in this review.
Finally, Appendix \ref{sec:TRHWH} provides a short introduction to 
Toeplitz operators and Wiener-Hopf factorisation, and is 
intended primarily to provide the necessary background results for readers who are not familiar with these topics,
while in Appendix \ref{sec:wardcan} we compare Ward's factorisation approach with the canonical Wiener-Hopf factorisation framework.
\\

%%%%%%%%%%%
\section{The gravitational field equations as an integrable system
\label{lsc}}
%%%%%%%%%%%%%

The field equations of gravitational theories in $D$ space-time dimensions are a system of non-linear partial differential equations (PDE's) for the space-time metric (and possibly other fields) for which
obtaining exact solutions is, in general, a difficult task. Exact solutions can, however, be obtained under simplifying assumptions, such as imposing spherical
symmetry. One well-known example of such a solution, which holds significant physical and mathematical interest, is 
the Schwarzschild solution.

Solutions to the gravitational field equations, also known as Einstein's field equations in the context of 
General Relativity, can be formulated in various coordinate systems.
Here, we focus on a specific subset of solutions to these field equations that exhibit a sufficient number of commuting isometries, enabling a dimensional reduction of the problem to two dimensions.
By performing a two-step dimensional reduction of the original theory (which we assume to have a vanishing cosmological constant), this leads to a system of non-linear, second-order PDEs \cite{Breitenlohner:1986um,Lu:2007jc,Nicolai:1991tt,Schwarz:1995af} that depend on 
two coordinates, which we take to be the Weyl coordinates $\rho, v$ (with $\rho > 0, v \in \mathbb{R}$):
\bea 
d \left( \rho \star A \right) = 0 \;\;\;,\;\;\; \text{with} \; \; A = M^{-1} dM  \;\;\;,\;\;\; M = M(\rho,v) \;,
\label{fi2d}
\eea
where $M \in G/H$ is a coset representative of the symmetric space $G/H$ that arises in the two-step reduction, and we assume $\det M=1$.
In \eqref{fi2d}, 
$\star$ denotes the Hodge star operator in two dimensions, satisfying
\bea
\star d \rho = - \lambda \, dv \;\;\;,\;\;\; \star dv = d \rho \;\;\;,\;\;\; (\star)^2 = - \lambda \, {\rm id} \;,
\label{stho}
\eea
where $\lambda = \pm 1$, depending on the form of the two-dimensional line element  
\bea
ds_2^2 = \sigma d\rho^2 + \varepsilon dv^2 \;,
\label{2dl}
\eea
with $\sigma, \varepsilon = \pm 1,  
\lambda = \sigma \varepsilon$. Note that $A$ satisfies
\bea
dA + A \wedge A = 0 \;.
\label{0cc}
\eea

The symmetric space
$G/H$ is invariant under an involution called {\it generalised transposition}, which we
denote by $\natural$, determined by the gravitational model under consideration. Hence $M^{\natural} = M$.
For $2 \times 2$ matrices, 
$\natural$ coincides with matrix transposition.
Examples of gravitational theories in $D$ dimensions and their associated 
symmetric spaces $G/H$  are:

\begin{itemize}

\item $D=4$ Einstein gravity: 

$G/H = SL(2, \mathbb{R})/SO(1,1)$ or $SL(2, \mathbb{R})/SO(2)$

\vskip 2mm

\item $D=4$ Einstein + Maxwell theory:  

$G/H = SU(2,1)/(SL(2, \mathbb{R}) \times U(1))$  or  $SU(2,1)/(SU(2)  \times U(1))$

\vskip 2mm

\item $D=5$ Einstein gravity: 

$G/H = SL(3, \mathbb{R})/SO(2,1)$ or $SL(3, \mathbb{R})/SO(3)$

\vskip 2mm

\end{itemize}

When $D=4$, the associated four-dimensional space-time metric takes the 
Weyl-Lewis-Papapetrou form,
\bea
ds_4^2 = - \lambda \, \Delta (dy  + B  d \phi  )^2 + \Delta^{-1} 
\left(  e^\psi \, ds_2^2 + \rho^2 d\phi^2 \right) \;,
\label{4dWLP}
\eea
where $\Delta (\rho,v), B(\rho,v)$ are determined  by the solution $M(\rho,v)$ of \eqref{fi2d}.
For instance, in the case of Einstein's field equations in vacuum, we have that
the matrix $M(\rho,v)$ is a symmetric $2 \times 2$ matrix of determinant $1$ given by \cite{Breitenlohner:1986um}
\bea
\label{M22}
M =
\begin{pmatrix}
 \Delta + {\tilde B}^2/\Delta &\;  {\tilde B}/ \Delta\\
{\tilde B}/\Delta  &\;  1/\Delta
\end{pmatrix} \;\;\;,\;\;\; \textrm{with} \;\;\;  \rho \star d {\tilde B} = \Delta^2 \, dB \;.
\eea
$\psi(\rho,v)$ is a scalar function
determined  by integration \cite{Lu:2007jc,Schwarz:1995af} from
\be
\partial_\rho \psi = \tfrac{1}{4} \, \rho \,  \Tr \left(A_\rho^2 - \lambda \, A_v^2\right) \,,\qquad \partial_v \psi = \tfrac{1}{2} \, \rho \, 
 \Tr \left(A_\rho A_v \right) \,. 
 \label{eq_psi}
\ee
Note that, in \eqref{eq_psi}, 
$\partial_v \left( \partial_\rho \psi \right) = \partial_\rho \left( \partial_v \psi \right)$, as can be verified by using \eqref{fi2d}  and \eqref{stho}. 

For examples of Weyl metrics in dimensions $D > 4$, see \cite{Emparan:2001wk}.

The non-linear field equations \eqref{fi2d} constitute an integrable system \cite{BabelonBernardTalon,Its}, i.e. they appear
as a compatibility condition for an associated linear system of PDE's, called a Lax pair, that involves,
besides the space-time coordinates $\rho, v$, a complex parameter $\tau$, called the {\it spectral parameter}.
The Lax pair takes the form \cite{Lu:2007jc}
\begin{equation}
	\tau \left( dX(\tau, \rho, v) + A (\rho,v)  X(\tau, \rho,v) \right) = \star \, dX (\tau, \rho,v) \;,
	\label{lax}
\end{equation}
where the spectral parameter $\tau$ varies on a spectral curve given by 
\bea
\omega = v + \frac{ \lambda}{2}  \, \rho \, \frac{\lambda - \tau^2}{\tau} \;.
\label{omt}
\eea
This linear system (for the unknown matrix $X(\tau, \rho, v)$) will be called the Breitenlohner-Maison linear system \cite{Breitenlohner:1986um}.\footnote{We present it here in the form given in \cite{Lu:2007jc}. Other equivalent linear systems may be considered, see \cite{Nicolai:1991tt,Figueras:2009mc,Katsimpouri:2012ky} for
a detailed analysis of their mutual relations.}
Omitting, for simplicity, the dependence on the spectral parameter and
the Weyl coordinates, the Lax pair \eqref{lax} becomes
\begin{equation}
	\tau \left( dX  + A  X \right) = \star \, dX  \;.
    \label{laxx}
\end{equation}

The complex parameter $\tau$ plays a crucial role here in allowing to bring in the tools of complex analysis 
and in reformulating the problem of obtaining solutions to \eqref{fi2d} in terms of a Riemann-Hilbert problem, as we shall explain in Section \ref{sec:RHP}.

Note that the spectral relation \eqref{omt} is invariant under the involution
\bea
i_{\lambda} (\tau) = - \frac{\lambda}{\tau}\;\;\;,\;\;\; \tau \in \mathbb{C} \backslash \{0\} \;.
\label{ilt}
\eea
As a consequence, the relation \eqref{omt} will be satisfied if we replace $\tau$ by any of the functions
\bea
\varphi_{\omega}^{\pm} (\rho, v)= \frac{-\lambda  (\omega-v) \pm \sqrt{(\omega-v)^2+\lambda \rho^2} }{\rho} \;,
\label{vpp}
\eea 
where $\varphi_{\omega}^- = - \lambda/ \varphi_{\omega}^+$, as can be easily verified. We denote the class of all functions of the form \eqref{vpp} by $\mathcal{T}$.

The integrability of the non-linear field equations \eqref{fi2d} can thus be expressed as follows, using the notation defined above.

{\theorem 
\cite{Lu:2007jc}
{\cite [Theorem 4.2]{Aniceto:2019rhg}}
Let $\varphi \in {\cal T}$. If, for a given $A = M^{-1} dM$, there exists $X(\tau, \rho, v)$ such that upon substituting $\tau = \varphi$, we have $X \in C^2, X^{-1} \in C^1$ and 
\bea
\tau \left( dX + A X \right) = \star \, dX \;, 
\label{varXA}
\eea
then $M$ is a solution to \eqref{fi2d}.\\
}

Note, however, that although the Breitenlohner–Maison linear system \eqref{laxx} is formally a system of linear PDE's for the unknown 
$X$, with coefficient $A(\rho, v)$, it should in fact be viewed as a system of equations for the pair of unknowns 
$(X,A)$. This is because $A$ is not given a priori;
rather, our goal is to determine a 1-form $A = M^{-1} d M$ such that \eqref{fi2d} holds.

 In the next
section we will review how this can be achieved by solving a Riemann-Hilbert problem which  yields both a solution
$M(\rho,v)$ to the field equations \eqref{fi2d} and a solution $X$
to the Breitenlohner–Maison linear system \eqref{laxx} with input $A = M^{-1} d M$.

%%%%%%%%%%
\section{{Solving the field equations by canonical Wiener-Hopf factorisation \label{sec:RHP}}}
%%%%%%%%%%%%%%%

We begin by briefly reviewing the Riemann-Hilbert (RH) problems that we will consider here, along with the related concept of Wiener-Hopf factorisation.

Let 
$\Gamma \subset \mathbb{C}$ be a closed contour, i.e., a simple closed path in the complex $\tau$-plane \cite{Rudin},
and let ${\cal M} (\tau)$ be an $n \times n$ matrix function, which we assume to be H\"older continuous \cite{MP,CG}, defined on $\Gamma$.
A RH problem with coefficient $\cal M$ consists in finding matrix valued (or vector valued) $\Phi_+$ and $\Phi_-$, belonging to certain classes
of functions analytic in the interior and the exterior of $\Gamma$, respectively, such that 
\bea
{\cal M} (\tau) \, \Phi_+ (\tau) = \Phi_- (\tau) \quad \textrm{on} \quad \Gamma \;.
\eea
This can be a {\it matricial, vectorial or scalar problem.}

A particularly important type of RH problems arises in the context of matrix factorisation.
By a Wiener-Hopf (WH) factorisation of a matrix ${\cal M}(\tau)$ as above, we mean a representation of the form
\bea
{\cal M} = M_- D M_+  \quad \textrm{on} \quad \Gamma \;,
\label{MmDMp}
\eea
where $M_+^{\pm 1}$ are analytic and bounded in the interior of $\Gamma$ (denoted $\textrm{int} \,  \Gamma$), 
$M_-^{\pm 1}$ are analytic and bounded in the exterior of $\Gamma$ (denoted $\textrm{ext} \, \Gamma$), and $D$ is a diagonal matrix of the form $D = \textrm{diag} (\tau^{k_1}, \dots, \tau^{k_n})$, with $k_i \in \mathbb{Z}$. It is well known \cite{MP} that any invertible H\"older continuous 
matrix admits a factorisation of the form \eqref{MmDMp}. If $k_j=0$  ($j=1, \dots, n$), then we have
\bea
{\cal M}= {M}_- {M}_+ \;,
\label{whf}
\eea
which is called a {\it canonical WH factorisation.} Note that, since, for any invertible constant matrix $K$, we have that
${\cal M}= \left( M_- K \right) \left( K^{-1} M_+ \right)$ is also a canonical WH factorisation, we can normalise
$M_+$ so that $M_+ (0) = \mathbb{I}_{n \times n}$, where $\mathbb{I}_{n \times n}$ denotes the identity matrix.

Clearly, obtaining such a factorisation is equivalent to solving a particular RH problem of the form
\bea
{\cal M} M_+^{-1} = M_- \;.
\label{mmmmp}
\eea

{\remark \label{remcont}

Note that, for a RH problem to be well formulated, one must specify both the class of analytic functions in which the solutions are sought, and the closed contour $\Gamma$ with respect to which the RH problem is formulated. Consequently,  RH problems with the same coefficient ${\cal M} (\tau)$ may have different solutions, depending on the choice of the solution class and the closed contour.\\

}

Here we will be considering  {\it admissible contours}, defined as simple closed paths encircling the origin
and invariant under the involution $i_{\lambda} (\tau) = - \frac{\lambda}{\tau}$.
We will also only be considering 
matrices that depend on a particular combination of the complex variable $\tau$ and the Weyl coordinates $\rho, v$, 
resulting from the composition of a matrix function ${\cal M} (\omega)$ with the spectral relation \eqref{omt}, i.e. of the form
\bea
{\cal M}_{\rho, v} (\tau) = {\cal M} \left( v + \frac{\lambda}{2}    \, \rho \, \frac{\lambda -   \tau^2}{\tau}  \right) \;,
\label{mont}
\eea
where $\rho,v$ play the role of parameters as far as the RH problem is concerned. We assume moreover that $\det {\cal M} (\omega) = 1 $ and 
${\cal M}^{\natural} (\omega) = {\cal M} (\omega)$. Such matrices are called {\it monodromy matrices.} They play the role of ``patching matrices" in Ward's construction.\footnote{The name monodromy matrix has its origin in the relationship of the theory of isomonodromic deformations with Einstein's field equations \cite{Korotkin:1994au,Korotkin:2023lrg}.}

Canonical Wiener-Hopf factorisation of monodromy matrices with respect to an admissible contour provides a method for obtaining a pair $(X,A)$, where $X$ is a solution to the Breitenlohner-Maison linear system \eqref{laxx} with input $A$, and $A$ is a solution to the non-linear field equations \eqref{fi2d}.
The following result was obtained in \cite{Aniceto:2019rhg}, and it holds under very general conditions. Here we present it in a slightly simplified form (omitting certain standard differentiability assumptions with respect to  $\rho$ and $v$; for further details, see \cite{Aniceto:2019rhg}).

{
\theorem
\label{theoraccr}
{\cite [Theorem 6.1]{Aniceto:2019rhg}}
Let $\Gamma$ be an admissible contour. Let ${\cal M}_{\rho,v} (\tau)$ be analytic, as well as its inverse, in a neighbourhood of $\Gamma$, and admitting a canonical WH factorisation with respect to $\Gamma$,
\bea
{\cal M}_{\rho, v} (\tau) = \left( M_{\rho, v}\right)_- (\tau) \, \left( M_{\rho, v}\right)_+ (\tau)\;\;\;\;\;\; 
\textrm{on} \; \;  \Gamma \;,
\label{whc}
\eea
where
\bea
\left( M_{\rho, v}\right)_+ (0) = \mathbb{I}_{n \times n} \;\;\; \forall \rho, v \;.
\label{normMp}
\eea
Then
\bea
\displaystyle{\lim_{\tau \rightarrow \infty}} \left( M_{\rho, v}\right)_- (\tau)  = M(\rho,v) 
\label{MmMrv}
\eea
satisfies the field equations \eqref{fi2d} and $X(\tau, \rho, v) := \left( M_{\rho, v}\right)_+ (\tau) $, with
$\tau = \varphi \in \mathcal{T}$, is a corresponding solution to the linear system \eqref{laxx}.\\
}

{\remark
\label{rem:xmx}

Note that, since ${\cal M}^{\natural} (\omega) = {\cal M} (\omega)$ and ${\cal M}_{\rho, v} (\tau)$
is obtained as in \eqref{mont}, one can prove \cite{Breitenlohner:1986um,Katsimpouri:2012ky,Camara:2017hez} that, if ${\cal M}_{\rho, v} (\tau)$ admits a canonical WH factorisation of the form \eqref{whc}, 
with $\left( M_{\rho, v}\right)_+ (\tau)$ normalised as in \eqref{normMp} and denoted $X(\tau, \rho,v)$, then one can rewrite the factor $ \left( M_{\rho, v} \right)_- (\tau)$ in such a way that the
canonical factorisation becomes 
\bea
{\cal M}_{\rho, v} (\tau) = X^{\natural} \left(- \frac{\lambda}{\tau}, \rho, v \right) M(\rho,v) X\left( \tau, \rho,v \right) \;.
\label{xmx}
\eea

Note that {\it different solutions} may be obtained from the {\it same monodromy matrix} if one chooses {\it different admissible contours $\Gamma$} with respect to which the factorisation is performed .
Therefore the monodromy matrix does not contain all the information regarding the solutions; there is an added information in the chosen contour, in line with what we mentioned in Remark \ref{remcont}.

Consider the following example.

{
\example

Take the Schwarzschild monodromy matrix, obtained from 
\be
\mathcal{M} (\omega ) = \begin{bmatrix}
\lambda \frac{\omega -m}{\omega +m} & 0 \\
	0 & \lambda  \frac{\omega+m}{\omega-m}
\end{bmatrix} \;\;\;,\;\;\; m \in \mathbb{R}^+ \;,
\label{monschwarz}
\ee
by composition with the spectral relation \eqref{omt} with $\lambda =1$  (we refer to \cite{Aniceto:2019rhg}
for a detailed discussion of the two cases $\lambda = \pm 1$).
 The resulting monodromy matrix is
 \be
\mathcal{M}_{\rho,v} (\tau)
= \begin{bmatrix}
	  \frac{\left(\tau - \tau_1\right) \left(\tau + 1/\tau_1 \right) }{\left(\tau - \tau_2\right) \left(\tau + 1/\tau_2 \right)} 
 & 0 \\
	0 & 
	   \frac{\left(\tau - \tau_2\right) \left(\tau + 1/\tau_2 \right) }{\left(\tau - \tau_1\right) \left(\tau + 1/\tau_1 \right)} 
	 \end{bmatrix} \;\;\;,\;\;\; 
\ee
where
\be
\tau_1 = \tau_1(\rho,v) =  \frac{v-m - \sqrt{\left(v-m\right)^2 + \rho^2}}{\rho} \,, 
\quad \tau_2 = \tau_2(\rho,v)=   \frac{  v+m  - \sqrt{\left(v+m\right)^2 + \rho^2} }{\rho}\,.
\label{tt12}
\ee
There are four possible classes
of admissible contours from which $\Gamma$ can be chosen, depending on which of the points $\tau_1$ and $\tau_2$ are inside or outside $\Gamma$.
Note that, since $\Gamma$ must be invariant under the involution $\tau \mapsto -1/\tau$, it has two fixed points, $\pm i$, and if $\tau \in \textrm{int} \,  \Gamma$ then $-1/\tau \in \textrm{ext} \, \Gamma$. Moreover, $\mathcal{M}_{\rho,v}$ must be continuous on $\Gamma$, so none of the points $\tau_i, -1/\tau_i$ $(i=1,2)$ belong to 
$\Gamma$.

The four types of admissible contours are depicted in Figure \ref{contout} for the case when $-m < v < m$, in which case we have that
$\tau_1 < -1 < \tau_2  < 0 $.

\begin{figure}[hbt!]
	\centering
	\includegraphics[scale=0.4]{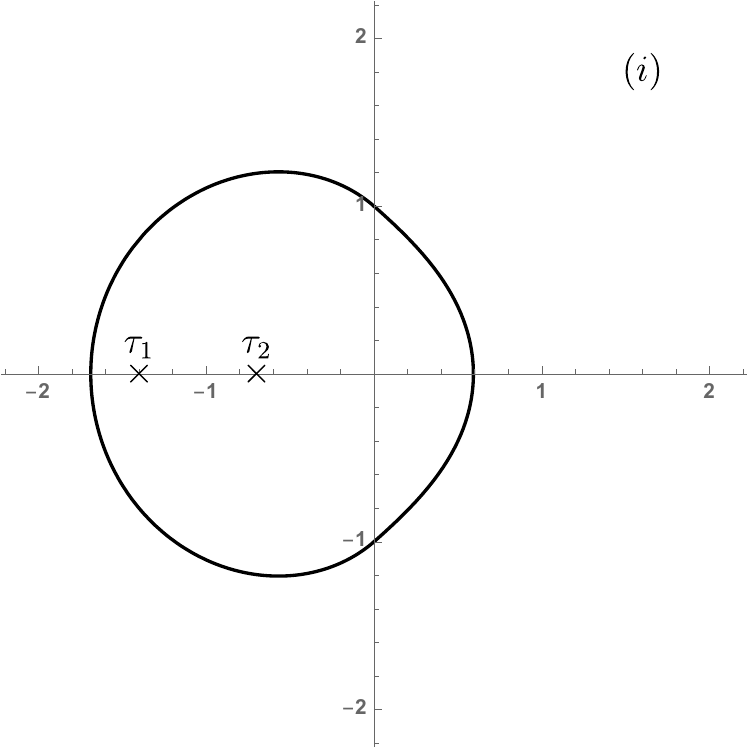}
	\includegraphics[scale=0.4]{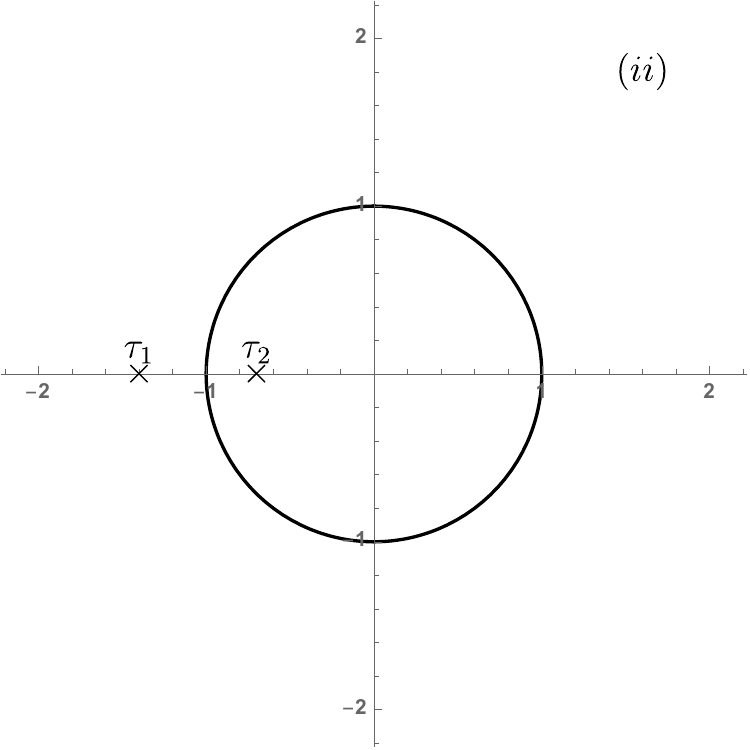}
	\includegraphics[scale=0.4]{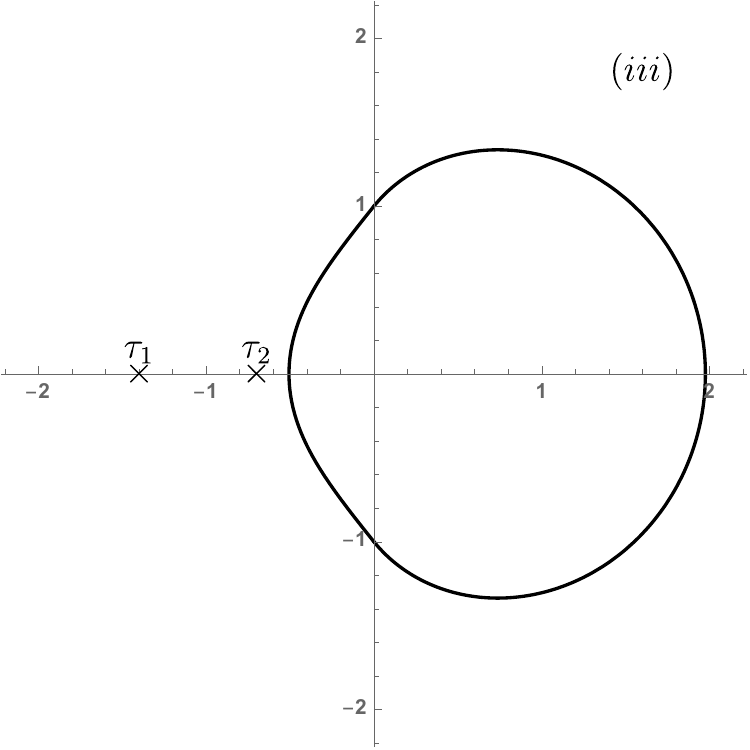}
	\includegraphics[scale=0.4]{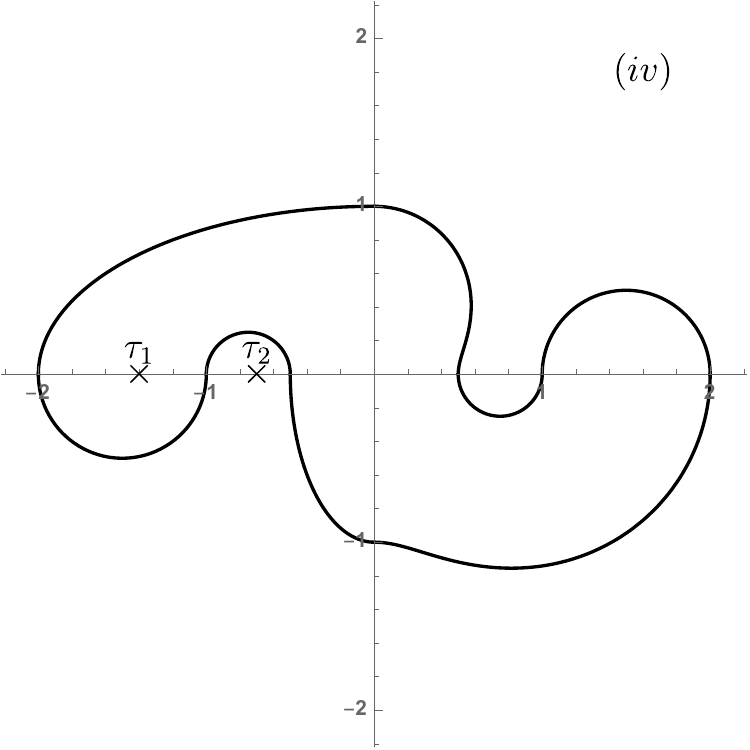}
		\caption{$-m < v < m$: four distinct choices of contours.
		\label{contout}}
\end{figure}

Factorising with respect to $\Gamma$ we obtain, in each  case, a canonical WH factorisation,
\be
\frac{\left(\tau - \tau_1\right) \left(\tau + 1/\tau_1 \right) }{\left(\tau - \tau_2\right) \left(\tau + 1/\tau_2 \right)}  = m_- (\tau) \, m_+ (\tau) \;,
\ee
which yields
\be
M (\rho,v)
= \begin{bmatrix}
	  m_- (\infty) \; & \;  0 \\
	0 \;  & \;
	 m_-^{-1} (\infty) 
	 \end{bmatrix}  =  
	 \begin{bmatrix}
	\Delta \; & \;  0 \\
	0 \;  & \;
	\Delta^{-1}  
	 \end{bmatrix} \;.
\ee

\noindent
(i) \underline{Case 1:} 
\be
m_+(\tau) = \frac{\tau_1}{\tau_2} \, \frac{\tau + 1 / \tau_1}{\tau + 1/ \tau_2} \;\;\;,\;\;\; m_-(\tau) = \frac{\tau_2}{\tau_1} \, \frac{\tau -  \tau_1}{\tau -  \tau_2} 
\;\;\;,\;\;\; \Delta =  \tau_2/\tau_1 \;\;\;,
\ee
yielding a solution which, under the change of coordinates
\bea
\rho =
\sqrt{r^2 - 2m r }  \, \sin \theta \;\;\;,\;\;\; v = (r -m) \cos \theta \;,
\label{sphext}
\eea
results in the four-dimensional line element
\begin{equation}
ds_4^2 = - \left(1 - \frac{2m}{r} \right) dt^2 + \left(1 - \frac{2m}{r} \right)^{-1}  dr^2 + r^2 \left( d \theta^2 +  \sin^2 \theta \, d \phi^2 \right) \:.
\label{schwarzmet}
\end{equation}
Thus, the solution  describes the {\it exterior region of the Schwarzschild black hole}.
\\

\noindent
(ii) \underline{Case 2:} 
\be
m_+(\tau) = \frac{1}{\tau_1 \, \tau_2} \, \frac{\tau - \tau_1}{\tau + 1 / \tau_2} \;\;\;,\;\;\; m_-(\tau) = \tau_1 \tau_2 \, 
\frac{ \tau + 1 /   \tau_1}{\tau -  \tau_2} 
\;\;\;,\;\;\; \Delta =  \tau_1 \tau_2  \;\;\;,
\ee
yielding a solution which, 
under the change of coordinates 
\bea
\rho =
\sqrt{2m \varrho - \varrho^2 }  \, \sinh \vartheta \;\;\;,\;\;\; v = (\varrho -m) \cosh \vartheta \;,
\eea
with $\varrho \in ]0, 2m[, \, \vartheta \in ]0, \infty[$, yields the four-dimensional line element
\begin{equation}
ds_4^2 =  \left(1 - \frac{2m}{\varrho} \right) dt^2-  \left(1 - \frac{2m}{\varrho} \right)^{-1}  d\varrho^2 + \varrho^2 \left( d \vartheta^2 + \sinh^2 \vartheta \, d \phi^2 \right) \:,
\label{schwarzmet2}
\end{equation}
which is an example of an $AII$-metric \cite{Griffiths:2009dfa}.\\

\noindent
(iii) \underline{Case 3:} 
\be
m_+(\tau) = \frac{\tau_2}{\tau_1 } \, \frac{\tau - \tau_1}{\tau - \tau_2} \;\;\;,\;\;\; m_-(\tau) =\frac{\tau_1}{ \tau_2} \, 
\frac{ \tau + 1 /   \tau_1}{\tau + 1/ \tau_2} 
\;\;\;,\;\;\; \Delta =  \tau_1 / \tau_2  \;\;\;,
\ee
yielding a solution which, 
under the change of coordinates 
\bea
\rho =
\sqrt{r^2 + 2m r }  \, \sin \theta \;\;\;,\;\;\; v = (r +m) \cos \theta \;,
\eea
results in the four-dimensional line element
\begin{equation}
ds_4^2 = - \left(1 + \frac{2m}{r} \right) dt^2 + \left(1 + \frac{2m}{r} \right)^{-1}  dr^2 + r^2 \left( d \theta^2 +  \sin^2 \theta \, d \phi^2 \right) \:.
\label{schwarzmetneg}
\end{equation}
This is the `negative mass' Schwarzschild solution \cite{Griffiths:2009dfa}, which has a naked curvature singularity at $r=0$.
\\

\noindent
(iv) \underline{Case 4:} 
\be
m_+(\tau) = \tau_1 \tau_2 \, \frac{\tau + 1 / \tau_1}{\tau -  \tau_2} \;\;\;,\;\;\; m_-(\tau) = \frac{1}{\tau_1 \tau_2}  \, \frac{\tau -  \tau_1}{\tau + 1/ \tau_2} 
\;\;\;,\;\;\; \Delta = \frac{ 1 }{\tau_1 \tau_2} \;\;\;.
\ee
This is analogous to Case 2 and yields the same solution upon using the transformation $v \mapsto -v$, which induces
the transformation $\tau_1 \tau_2 \mapsto 1/(\tau_1 \tau_2)$.

}

%%%%%%

\section{Canonical Wiener-Hopf factorisation: the question of existence \label{sec:bwh}}

Theorem \ref{theoraccr} presupposes the existence of a canonical WH factorisation of the monodromy matrix ${\cal M}_{\rho,v} (\tau)$. But does such a factorisation actually exist?  This is a natural and important question, raised by several authors, see for example \cite{Ward1982, Breitenlohner:1986um, Katsimpouri:2012ky}.

Indeed, for any invertible H\"older continuous matrix function, a WH factorisation of the form 
\eqref{MmDMp} always exists,   but, in general, it will not be canonical, i.e. we do not have $k_i=0$ for all $i$.
Even in the case of scalar functions, the factorisation is canonical only if the winding number of the function around the origin is zero \cite{GK,CG,MP}; for example, a simple function like $f(\tau) = \tau$ does not admit a canonical WH factorisation. 
Therefore, what follows is a surprising result: although simple in form, it reveals that functions depending on $\tau$ via composition with the spectral relation, as in \eqref{mont},  have very special properties.

{\theorem \cite{Cardoso:2017cgi} \label{theo:joh}
Let $g$ be a scalar H\"older continuous function on an admissible contour $\Gamma$ and let $g_{\rho,v} (\tau) = g\left( v + \frac{\lambda}{2} \rho \frac{\lambda - \tau^2}{\tau} \right)$ be non-vanishing on $\Gamma$. Then $g_{\rho,v} (\tau)$ admits a canonical factorisation.
\\

}

The canonical factorisation of $g_{\rho,v} (\tau) $ in the above theorem can be explicitly obtained using the complementary projections
\bea
P_{\Gamma}^{\pm} = \frac12 \left( \mathrm{Id} \pm S_{\Gamma} \right) \;,
\label{proje}
\eea
where $ \mathrm{Id}$ denotes the identity operator and $S_{\Gamma}$ is the singular integral operator with Cauchy kernel defined by (see Appendix \ref{sec:TRHWH})
\bea
  \left( S_{\Gamma} \varphi \right) (\tau) = \frac{1}{ \pi i} \mathrm{P.V.}  \int_{\Gamma} \, \frac{\varphi(u)}{u - \tau} \, du  \;\;\;,\;\;\; \tau \in \Gamma \;.
 \label{Ppm}
 \eea
 $S_{\Gamma}$ maps  the space $C^{\mu} _{\Gamma}$ of all H\"older continuous functions (on $\Gamma$) with exponent $\mu \in \,  ] 0, 1[$ onto itself \cite{MP, CG}. We  then have 
\bea
g_{\rho,v} (\tau) = \left( g_{\rho,v}\right)_-  (\tau) \, \left( g_{\rho,v} \right)_+ (\tau)
\label{gfpm}
\eea
with 
\bea
\left(g_{\rho,v} \right)_{\pm}  = e^{ P_{\Gamma} ^{\pm}  \log g_{\rho,v} } \;.
\label{gpm}
\eea

Theorem \ref{theo:joh}  has significant consequences, in particular the following, which is a consequence of 
well known results in the theory of WH factorisation {\cite [Chapter 4] {CG}}, \cite{LS, MP}.

{\corollary
\label{cor:iii}
Let $\Gamma$ be an admissible contour and let ${\cal M}_{\rho,v} (\tau)$, defined as in \eqref{mont}, be in $\left( C^{\mu}_{\Gamma} \right)^{n \times n}$, with non-vanishing determinant on $\Gamma$.

(i) If ${\cal M}_{\rho,v} (\tau)$ is diagonal, then it admits a canonical WH factorisation that can be explicitly obtained by factorising each diagonal element according to \eqref{gfpm} and \eqref{gpm}.

(ii)  If ${\cal M}_{\rho,v} (\tau)$ can be reduced to triangular form by left multiplication with a matrix that is analytic and bounded in $\mathrm{ext} \, \Gamma$, along with its inverse, and  right multiplication with a matrix that is analytic and bounded in $\mathrm{int} \, \Gamma$, along with its inverse, then
${\cal M}_{\rho,v} (\tau)$  admits a canonical factorisation. \\}

For matrix functions which do not satisfy the conditions of Corollary \ref{cor:iii}, the question of whether they admit a canonical factorisation may be considerably more complex. To address it, we will use the close relationship between WH factorisation and the study of Toeplitz operators (see Appendix \ref{sec:TRHWH} for a very brief introduction to this topic). The latter are compressions of multiplicative operators into the Hardy space $H^2_+ = P_{\Gamma}^+ L^2 (\Gamma)$ (where $L^2 (\Gamma)$ denotes the space of all square-integrable functions on $\Gamma$), or its vectorial analogues $\left( H^2_+ \right)^n$ in the matricial case. Concretely, given an $n \times n$ matrix function $G$ whose elements are bounded functions on $\Gamma$,
the Toeplitz operator $T_G$ is defined by
\bea
T_G = P_{\Gamma}^+ G P_{\Gamma}^+\vert_{(H^2_+)^n} : (H^2_+)^n \rightarrow (H^2_+)^n \;,
\label{toepG}
\eea
where $(H^2_+)^n $ is the space of $n \times 1$ vectorial functions with elements in $H^2_+$ and $P_{\Gamma}^+$ is
the projection defined in \eqref{proje}, applied componentwise. $G$ is called the {\it symbol} of the Toeplitz operator. The connection with WH factorisation comes from the following results presented in more detail in Appendix  \ref{sec:TRHWH}.

{\theorem  \label{theo:G}
If $G$ is an invertible H\"older continuous matrix function defined on $\Gamma$, then $G$ admits a canonical WH factorisation if and only if the Toeplitz operator with symbol $G$, defined in \eqref{toepG}, is invertible.\\
}

{\theorem 
With the same assumptions as in Theorem \ref{theo:G}, if $\det G = 1$ then $T_G$ is invertible in 
$\left( H_+^2 \right)^n$ if and only if it is injective.\\

}

Now, it follows from the definition of a Toeplitz operator and from the complementarity of the projections $P_{\Gamma}^+$ and $P_{\Gamma}^-$ defined in  \eqref{proje} that $T_G$
is injective (i.e. $\ker T_G = \{ 0 \}$)  if and only if the only solution to the vectorial RH problem 
\bea
G \Psi_+ = \Psi_- \;\;\;,\;\;\; \Psi_{\pm} \in \left( H^2_{\pm} \right)^n 
\label{Gpsipm}
\eea
is the trivial solution $\Psi_\pm =0$, where 
\bea
H_{\pm}^2 = P_{\Gamma}^{\pm} L^2 (\Gamma) \;.
\eea
We call the RH problem \eqref{Gpsipm} {\it the injectivity RH problem} for the Toeplitz operator $T_G$.  Thus, we have the following.

{\proposition
With the same assumptions as in Theorem \ref{theo:G}, with $\det G =1$, $G$ has a canonical factorisation if and only if \eqref{Gpsipm} admits only the trivial solution.\\
}

Note that although \eqref{Gpsipm} and the RH factorisation problem \eqref{mmmmp} have a similar form, they are in fact quite different, since in \eqref{Gpsipm} one looks for vectorial solutions in $\left( H_{\pm}^2 \right)^n$, in particular with $\Psi_- (\infty) =0$, while in \eqref{mmmmp} the solutions are sought in the space of $n \times n$ matrix functions which are analytic and bounded either in the exterior or in the interior of $\Gamma$ (see Remark \ref{remcont}). Indeed,  these two RH problems address different questions: by solving \eqref{Gpsipm} we determine the kernel of $T_G$ and answer the question of whether or not there exists a canonical WH factorisation, while the solution to \eqref{mmmmp} provides that factorisation, if it exists.

Also note that the existence of a canonical factorisation is stable under small perturbations, as 
is well known (see Appendix \ref{sec:TRHWH}).

For rational matrices, it is, in principle, always possible to determine whether the associated injectivity  RH problem admits only the zero solution. In practise, however, this can be a highly nontrivial task. It is important to note that the known results for 
 general rational matrices \cite{CG,MP,LS} cannot be directly applied in our setting, due to the specific dependence of the monodromy matrices on $\tau$, as made evident  by Theorem \ref{theo:joh} and Corollary \ref{cor:iii}.

However, for $2 \times 2$ rational matrices obtained as in \eqref{mont} from
\bea
{\cal M} (\omega) = \frac{1}{q (\omega) }  \begin{pmatrix}
p_{11} (\omega) & p_{12} (\omega)  \\
p_{12} (\omega) & p_{22} (\omega)
\end{pmatrix},
\label{pomeg}
\eea
where $q, p_{11}, p_{12}, p_{22}$ denote polynomials of degree $n, k_{11}, k_{12}, k_{22}$, respectively, 
and where we assume that $q$ only has simple zeroes, 
and furthermore denoting
\bea
N_1 = \max \{k_{11}, k_{12} \} \;\;\;,\;\;\; N_2 = \max \{k_{12}, k_{22} \} ,
\label{NNkk}
\eea
we have the following simple criteria.

{
\theorem \cite[Theorem 3.5] {Camara:2024ham} 
\label{theoNNn}
With the notation above and  ${\cal M}_{\rho,v} (\tau)$
given by \eqref{mont}, let $\Gamma$ be an admissible contour where $\det {\cal M}_{\rho,v} (\tau) = 1$.
Then we have:
\\
(i) if $N_1 + N_2 < 2n$, then ${\cal M}_{\rho,v} (\tau)$ has a canonical WH factorisation w.r.t. $\Gamma$, for all $\rho,v$;\\
(ii) if $N_1 + N_2 = 2n$, then ${\cal M}_{\rho,v} (\tau)$ has a canonical WH factorisation 
w.r.t. $\Gamma$ for all $\rho,v$, except for the points on a certain curve $\cal C$ in the Weyl plane;\\\
(iii) the case where  $N_1 + N_2 > 2n$ can be reduced to case (ii). \\

}

To exemplify this result and illustrate how the curve $\cal C$, along which the canonical WH factorisation of ${\cal M}_{\rho,v} (\tau)$ breaks down, is determined, we consider the following.

{\example The Kerr black hole (with $\lambda = 1$)

Consider the non-extremal Kerr matrix (see \cite{Woodhouse1988,Woodhouse1997,Katsimpouri:2012ky})
\begin{equation}
{\cal M}(\omega)= \frac{1}{\omega^2 - c^2} \begin{pmatrix}
(\omega - m)^2 + a^2 & \quad 2 a m   \\
2 a m  & \quad (\omega + m)^2 + a^2 
\end{pmatrix} \;\;\;,\;\;\; c = \sqrt{m^2 - a^2} > 0 .
\label{Monokerr}
\end{equation}
The corresponding matrix ${\cal M}_{\rho,v} (\tau)$ is given by
\bea
{\cal M}_{\rho,v} (\tau) = \frac{\tau^2}{q_{4} (\tau) } \,  \tilde{\cal M}_{\rho,v} (\tau) \;,
\label{calm22}
\eea
where 
\bea
q_4 (\tau) &=& \frac14 \left[ \rho^2 (1 - \tau^2)^2 + 4 (v^2 - c^2) \tau^2 + 4  \rho v \tau (1-\tau^2) \right]  \;,
\nonumber\\
\tilde{\cal M}_{\rho,v} (\tau) &=&  \begin{pmatrix}
 (v - m +  \rho \, \frac{(1- \tau^2)}{ 2 \tau } )^2 + a^2 &  2 a m  \\ 
 2 a m & (v + m +  \rho \, \frac{ (1- \tau^2)}{ 2 \tau} )^2 + a^2
\end{pmatrix} .
\label{qMn2}
\eea
Note that this corresponds to the case considered in Theorem \ref{theoNNn} $(ii)$ , with  $n=2, 
k_{11} = k_{22} = 2, k_{12} = 0, 
N_1 = N_2 = 2, 2n = 4 = N_1 + N_2$. Now let
\bea
\tau_1 = \frac{v - c - \sqrt{(v-c)^2+ \rho^2} }{\rho} \;\;\;,\;\;\;
\tau_2 = \frac{v + c - \sqrt{(v+c)^2+ \rho^2} }{\rho} \;.
\label{vpp2}
\eea 
Solving the injectivity RH problem \eqref{Gpsipm} by means of a generalisation of Liouville's theorem \cite[Lemma, Section 3]{Camara:2017hez} one obtains an algebraic homogeneous linear system of four equations for four unknown constants, that can be solved by Cramer's rule; the determinant $D(\rho,v)$ of that linear system must be different from zero for the only solution to be zero (see \cite{Camara:2024ham} for details on the calculation). Thus 
${\cal M}_{\rho,v} (\tau)$ will have a canonical factorisation if and only if $(\rho,v)$  does not lie on the curve defined by $D(\rho,v) = 0$, which can be shown in this case to be equivalent to
\bea
&&- 16  (m - v)^2 \tau_1^2 \tau_2^2 + \rho^2 \left( 1 + 4 \tau_1^3 \tau_2 + 6 \tau_1^2 \tau_2^2 + 4 \tau_1 \tau_2^3 + \tau_1^4 \tau_2^4 \right)\nonumber\\
&& \qquad \qquad  - 8 \rho (m-v) \tau_1 \tau_2 \left( - \tau_1 - \tau_2 + \tau_1^2 \tau_2 + \tau_1 \tau_2^2 \right) = 0 \;.
\label{funcfg}
\eea
This condition describes a curve $\mathcal{C}$ in the Weyl half-plane of the coordinates $\rho > 0, v \in \mathbb{R}$. If we express \eqref{funcfg} in terms of prolate spheroidal coordinates $(u,y)$
 (see \cite{Harmark:2004rm,Katsimpouri:2012ky}),
\bea
v = u \, y \;\;\;,\;\;\; \rho= \sqrt{(u^2 - c^2)(1-y^2)},
\eea
where
\bea
c < u < + \infty \;\;\;,\;\;\; |y| < 1 \;,
\eea
we obtain from \eqref{funcfg} 
\bea
u (y) = \sqrt{m^2 - a^2 y^2 } \;.
\eea
This equation defines the ergosurface of the non-extremal Kerr black hole in four dimensions, in the exterior region, where the metric component $g_{tt}$ of the Kerr metric vanishes.

The curve $\mathcal{C}$ defined by \eqref{funcfg} is represented in Figure \ref{ergos}, in the Weyl half-plane. The region between the curve  $\mathcal{C}$ and the axis $\rho = 0$ represents the ergosphere (the region between the ergosurface $\mathcal{C}$ and the outer horizon of the non-extremal Kerr black hole),
while the complementary region describes the region outside the ergosphere.

\begin{figure}[hbt!]
	\centering
	\includegraphics[scale=1]{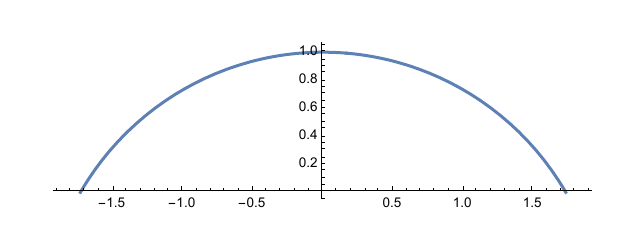} 
	\caption{Curve $\mathcal{C}$ in the Weyl coordinates upper half-plane $(\rho >0, v)$ for the values $m= 2, a = 1$. The horizontal axis represents $v \in \mathbb{R}$, while the vertical axis represents 
	$\rho> 0$.
	\label{ergos}}
\end{figure}

}

%%%%%%%%%%%

\section{Beyond Wiener-Hopf factorisation: $\tau$-invariance and solution generation by multiplication \label{sec:tauinv}}

The question of existence of a canonical factorisation was addressed in Section \ref{sec:bwh}. However, even when such a factorisation exists, two further questions naturally arise. The first concerns the explicit construction of a canonical WH factorisation. For matrix functions, this always depends on their specific form and requires the development of tailored methods for different classes of matrices \cite{Its}. Recent advances in this area
have made it possible to apply new factorisation techniques to a variety of matrix classes.
We present two examples in Section  \ref{sec:exam}.
The second question is whether any solution to \eqref{fi2d} can be obtained from a canonical WH factorisation of an associated monodromy matrix, as described in Section \ref{sec:bwh}. The answer to this second question is in the negative, 
as the following example demonstrates.

{
\example \label{ex:cosms}

Consider the cosmological Kasner solution to Einstein's field equations in four dimensions,
\bea
ds^2_4 = - dt^2 + \sum_{i=1}^3 \, t^{2 p_i} \, dx_i^2 \;\;\;,\;\;\; \sum_{i=1}^3 p_i = 1 \;\;\;,\;\;\;
\sum_{i=1}^3 p_i^2 = 1 \;,
\eea
with 
\bea
p_1 = p_3 = \frac23 \;\;\;,\;\;\; p_2 = - \frac13 \;,
\eea
which is of the form \eqref{4dWLP} with 
$\lambda =-1$. The corresponding  Weyl-Lewis-Papapetrou form of the line element is
\bea
ds_4^2 = \left( \frac{\rho}{2} \right)^4 \,  dy^2   +  \frac94 \left( \frac{\rho}{2} \right)^4 \left( dv^2 - d\rho^2 \right)  + 
\left( \frac{2}{\rho} \right)^2  d\phi^2  \;,
\label{linrv}
\eea
where
\bea
\rho = 2 t^{1/3} \;\;\;,\;\;\; v = \frac23 x_1 \;\;\;,\;\;\; y = x_3 \;\;\;,\;\;\; \phi = \frac12 x_2 \;.
\eea
In this case, we have that $M(\rho,v)$ actually only depends on $\rho$ \cite{Aniceto:2019rhg},
\bea
\label{Mkas}
M(\rho,v) = M(\rho) = \begin{pmatrix}
\left(\frac{\rho}{2} \right)^4 & 0 \\
0 & \left( \frac{\rho}{2} \right)^{-4}
\end{pmatrix} \;.
\eea
Due to the particularly simple form of $M(\rho,v)$, one can explicitly solve the Breitenlohner-Maison linear system \eqref{laxx} with coefficient $A = M^{-1} d M $, which results in the general solution
\bea
X(\tau, \rho, v) = \begin{pmatrix}
    \frac{\tau^2}{\rho^2} \, c_1 \quad & \frac{\tau^2}{\rho^2} \, c_2 \\
    \frac{\rho^2}{\tau^2} \, c_3 \quad &  \frac{\rho^2}{\tau^2} \, c_4
\end{pmatrix} \;,
\eea
where $c_i, i=1, \dots, 4$ are arbitrary integration constants. It is clear, on the one hand, that the matrix function $X$ cannot be normalised to the identity at $\tau =0$, so it cannot be a factor in a canonical WH factorisation. On the other hand, the product \eqref{xmx}, for any matrix of the form \eqref{Mkas}, is a constant matrix involving only the values of  $c_i, i=1, \dots, 4$, so in this case it is not possible to construct a meaningful monodromy matrix which would, by canonical WH factorisation, provide a solution to the linear system (see Remark \ref{rem:xmx}).\\
}

It is thus natural to look for new approaches - extending beyond, but still including, the Wiener–Hopf factorisation framework -  that allow to construct other classes of solutions. A motivation for generalising the Riemann-Hilbert method to solving \eqref{fi2d}, as  described in Theorem \ref{theoraccr},
 comes from the structure of  its proof presented in \cite{Aniceto:2019rhg}. That proof relies in a crucial way on showing that the factor $X$ in a canonical WH factorisation of ${\cal M}_{\rho,v} (\tau)$ satisfies a boundary value problem on $\Gamma$, which is not a Riemann-Hilbert problem (see eq. (6.17) in \cite{Aniceto:2019rhg}).
 This leads  to the conclusion that a certain expression involving $X$ and its derivatives, of the form
\bea
\label{GX}
G_{M,X} (\tau, \rho, v) &=&
 \tau \, d M +  \frac{1}{\rho}\left[(\lambda-\tau^2)d\rho + 2\lambda \tau dv\right] 
M \frac{\partial X}{\partial \tau} X^{-1} 
\nonumber\\	&&  
	+ \frac{\tau^2+\lambda}{\tau} M \left(\frac{\partial X}{\partial \rho} d\rho + \frac{\partial X}{\partial v} dv \right)X^{-1} \;,
    \eea
is independent of the spectral parameter $\tau$. Remarkably, this property, by itself and without
involving any prior factorisation of a monodromy matrix, leads to the desired conclusion, as stated in the  following result.

 {
\theorem
\label{theo3.3}
{\cite [Theorem 3.3] {Camara:2022gvc}}
Let $\Gamma$ be an admissible contour and let $X(\tau, \rho, v)$ and $M(\rho, v)$ be matrix functions such that $X^{\pm 1}$ are analytic with respect to $\tau$ in a neighbourhood of $\Gamma$ for all 
$(\rho, v)$, and $X^{\pm 1}, M$ are of class $C^2$ with respect to $(\rho, v)$, and $M = M^{\natural}$.
Let $G_{M,X} (\tau, \rho, v)$ be defined by \eqref{GX}, where we omitted the dependence of $X$ and $M$  on $(\tau, \rho, v)$ and $(\rho,v)$, respectively,
on the right hand side for simplicity.

If, for all $(\rho,v)$, 
\bea
\frac{\partial}{\partial \tau} G_{M,X} (\tau, \rho, v) = 0 \quad \text{on} \quad \Gamma,
\label{tderGt}
\eea
then $M(\rho,v)$ is a solution to the field equation \eqref{fi2d} and $X(\tau, \rho, v)$, with $\tau = \varphi \in {\cal T}$, is a solution to the linear system \eqref{laxx}.\\

}

We refer to \eqref{tderGt} as the property of $\tau$-invariance.

It is natural to ask how one can explicitly construct matrices $X$ and $M$ that satisfy the assumptions of Theorem \ref{theo3.3} and for which
\eqref{tderGt} holds. It follows from the proof of Theorem 3.3 in \cite{Aniceto:2019rhg} that any pair of functions $X(\tau, \rho,v)$ and $M(\rho,v)$, obtained from a canonical WH factorisation of a monodromy matrix, will do; but, just as there are no general methods to obtain the WH factorisation of a matrix function, there is also no systematic procedure to obtain functions $X$ and $M$ satisfying \eqref{tderGt}.

Nevertheless, by using the fact that, for any $\omega_i \in \mathbb{C}$ (with $\omega_i \neq v$), if $\tau_i, {\tilde \tau}_i = - \lambda/\tau_i$ are the two roots of the equation
\bea
\omega_i = v + \frac{\lambda}{2}  \, \rho \, \frac{\lambda - \tau^2}{\tau} \,,
\eea
we have 
\bea
\left(  \frac{\tau -  {\tilde \tau}_i}{\tau - \tau_i} \right)^{-1} 
= \frac{\tau_i}{ {\tilde \tau}_i} \; \; \frac{ - \frac{\lambda}{\tau} - {\tilde \tau}_i}{- \frac{\lambda}{\tau} - { \tau}_i}
\;,
\eea
we can characterise a class of functions satisfying the conditions of Theorem \ref{theo3.3}, which are not obtained from the canonical WH factorisation of a monodromy matrix. We have the following.

{\proposition \label{prop:RN} \cite[Proposition 3.6]{Camara:2022gvc}
Let $\omega_i, \tau_i, {\tilde \tau}_i$ be defined as above. Let
\bea
R_i (\tau, \rho, v) = \frac{\tau_i}{{\tilde \tau}_i} \frac{\tau - {\tilde \tau}_i}{\tau - \tau_i} \;\;\;,\;\;\; N_i(\rho, v) = 
\frac{{\tilde \tau}_i}{{ \tau}_i} = - \frac{\lambda}{\tau_i^2} = - \lambda {\tilde \tau}_i^2 \;\;\;,\;\;\; i= 1, \dots, n \;,
\eea
and 
\bea
R(\tau, \rho, v) &=& {\rm diag} \left( R_i^{\alpha} (\tau, \rho, v) 
\right)_{i=1, \dots, n} \;,
\nonumber\\
N(\rho,v) &=& {\rm diag} \left( N_i^{\alpha} ( \rho, v) 
\right)_{i=1, \dots, n} \;,
\label{RN}
\eea
with $\alpha \in \mathbb{R}$. Then $R(\tau, \rho, v) $ and $N(\rho,v)$ satisfy
\bea
\frac{\partial}{\partial \tau} G_{R,N} (\tau, \rho, v) = 0 
\eea
on any admissible contour $\Gamma$ such that $\tau_i, {\tilde \tau}_i  \notin \Gamma$ for all $i=1, \dots, n$.
\\
}

{
\corollary \cite[Corollary 3.7]{Camara:2022gvc}
With the same assumptions as in Theorem \ref{theo3.3}, 
$N(\rho,v)$ defined in \eqref{RN} yields a solution to the field equations \eqref{fi2d}.\\

}

Moreover, as we show next, we can use the functions
defined in Theorem \ref{theo3.3} to generate new families of 
solutions 
by multiplication, provided that certain commutation relations involving both the solutions to \eqref{fi2d} and the corresponding solutions to the linear system \eqref{laxx} hold. In what
follows we denote by ${\cal F}$ the class of all pairs
$(M, X)$ of matrices $X = X (\tau, \rho, v) , M= M(\rho,v)$ satisfying 
the assumptions of Theorem \ref{theo3.3} and such that \eqref{tderGt} holds.

We have the following.

{\theorem {\cite [Theorem 3.9] {Camara:2022gvc}}
\label{theoMN}

Let $\Gamma$ be an admissible contour and let
$(N,R) , (M,X) \in {\cal F}$.
Then, if $R$ commutes with $X$ and its derivatives with respect to $\tau, \rho, v$, $M$ commutes with $R$ and its derivatives with respect to 
$\tau, \rho, v$, and $N$ commutes with $X$ and its derivatives with respect to $\tau, \rho, v$, we have that $R X$ is a 
solution to the linear system \eqref{laxx}, with 
$\tau= \varphi \in \mathcal{T}$, for the coefficient $A = (M N )^{-1} d (M N)$, and $(M N) (\rho, v)$ is a solution to the field equations \eqref{fi2d}.
\\
}

{\remark \label{rem:diag}

The commutation relations in Theorem \ref{theoMN} hold, in particular, for diagonal matrices.\\}

As concrete examples of the solution generating method
presented in Theorem \ref{theoMN} we obtain first a cosmological Kasner
solution from the interior region of the  Schwarzschild solution
using matrices $N$ defined in Proposition \ref{prop:RN}, then the solution of Example  \ref{ex:cosms} and,  finally, we show how to
take advantage of some remarkable properties of Einstein-Rosen waves to obtain a family of deformations of a Kasner solution.

{ \example

Consider the Schwarzschild monodromy matrix obtained from \eqref{monschwarz}
by substituting $\omega = v + \frac{\lambda}{2} \rho \frac{\lambda - \tau^2}{\tau}$ with $\lambda = -1$.
Following \cite[Section 3.1]{Camara:2022gvc}, we consider the solution that describes the interior of the
Schwarzschild solution in the region of the Weyl plane delimited by $\rho = 0, - m < v < m ; \rho = m + v ; \rho = m - v$. This solution arises by canonically factorising ${\cal M}_{\rho,v} (\tau)$ with respect to a contour $\Gamma $ in the $\tau$-plane that passes through the fixed points $\tau = \pm 1 $ of the involution $\tau \mapsto 1/\tau$, with 
\bea
{ \tau}_1 = \frac{m - v + \sqrt{ (m-v)^2 - \rho^2}}{\rho} \;\;\;,\;\;\; { \tau}_2 = \frac{- m - v + \sqrt{ (m+v)^2 - \rho^2}}{\rho} 
\eea
inside $\Gamma$, so that ${\tilde \tau}_1 = 1/\tau_1, {\tilde \tau}_2 = 1/\tau_2$ lie outside $\Gamma$. We obtain, for the interior Schwarzschild solution,
\bea
M (\rho,v) = \begin{pmatrix}
- \frac{{ \tau}_2}{{ \tau}_1}
 & 0
 \\
 0 &- \frac{{ \tau}_1}{{ \tau}_2} 
  \end{pmatrix} \;.
  \eea
If we now choose, according to Proposition \ref{prop:RN}
\bea
N_1(\rho,v) &=& {\rm diag} \left( \left(  \frac{{\tilde \tau}_2}{{ \tau}_2} \right)^{1/2}, 
 \left( \frac{{\tau}_2}{{ \tilde \tau}_2} \right)^{1/2} \right) = {\rm diag} \left( -\frac{1}{\tau_2}, -\tau_2 
 \right) \;, \nonumber\\
 N_2(\rho,v) &=& {\rm diag} \left(  \tau_1, \frac{1}{\tau_1} \;,
 \right) 
\eea
and multiplying $M N_1 N_2$, according to Theorem \ref{theoMN}, we get
\bea
M_m (\rho,v) =  {\rm diag} \left( 1, 1 \right) \;.
\eea
The associated matrix one-form $A_m = M^{-1}_m d M_m$ vanishes, and the 
factor $\psi$, obtained by integrating \eqref{eq_psi}, is constant. We take it to be zero. The associated space-time metric \eqref{4dWLP}
takes the form 
 \bea
ds_4^2 = 
 -d \rho^2 + dv^2 +  \rho^2 d\phi^2 + dy^2 \;,
 \label{kasn010}
\eea
which describes a cosmological Kasner solution with exponents $(p_1, p_2, p_3) = (0, 1, 0)$.

}

{ \example \cite{Aniceto:2019rhg} \label{ex:kas}

Now consider the monodromy matrix 
\bea
{\cal M}_{\rho, v} (\tau) = 
 \begin{bmatrix}
	   \omega^4
		\; \; & \;\;	0 \\
		0\;  \; & \; \;	  \omega^{-4}
\end{bmatrix} _{\omega = v  + \frac12 \rho \frac{(1 + \tau^2)}{\tau} } \;,
\label{monN}
\eea
where in \eqref{omt} we took $\lambda = -1$. Its canonical factorisation with respect to an admissible contour $\Gamma$, with 
\bea
 { \tau}_0 (\rho,v)  = \frac{-  v  -  \sqrt{v^2 - \rho^2}}{\rho} 
 \label{t0til}
 \eea
 in  the exterior of $\Gamma$, is given by \eqref{xmx} with $X=X_c, M= M_c$, where
 \bea
 M_c (\rho, v) &=&  {\rm diag} \left( \left( \frac{\rho}{2} { \tau}_0  \right)^4,   \left( \frac{\rho}{2} { \tau}_0 \right)^{-4} \right) \;, \nonumber\\
 X_c (\tau, \rho, v) &=& \begin{pmatrix} 
 \left( \frac{\tau - \tau_0}{\tau_0} \right)^4 & 0 \\
 0 & \left( \frac{\tau_0}{\tau - \tau_0} \right)^4 
 \end{pmatrix} \;.
\label{Mac}
\eea
If we multiply \eqref{Mac}, similarly to the previous example, by
\bea
{\rm diag} \left( { \tau}_0^{-4},   { \tau}_0^{4} \right) \;,
\eea
we obtain the cosmological Kasner solution in \eqref{Mkas} of Example \ref{ex:cosms}. 
Note that $(M, X)$ satisfy the invariance property \eqref{tderGt} with 
\bea
X (\tau, \rho, v) &=& \begin{pmatrix} 
 \left( \tau^2 + 2 \frac{v}{\rho} \tau + 1  \right)^2 & 0 \\
 0 & \left(  \tau^2 + 2 \frac{v}{\rho} \tau + 1 \right)^{-2}
 \end{pmatrix} \;.
 \label{Xmer}
\eea
\\

}

{ \example 
 \underline{Einstein-Rosen waves:}

We consider the following family of diagonal matrices,
\bea
 {\cal M}(\omega) =
\begin{pmatrix}
e^{ 4 b \, e^{- a k}  \cos \left( k  \omega \right) }& 0 \\
0 & e^{ - 4 b \, e^{- a k}  \cos \left( k  \omega \right) }
\end{pmatrix} \;,
\label{monpesuper}
\eea
where  $a, b \in \mathbb{R}, a >0$ are constants, and where the parameter $k$ takes values in $\mathbb{R}_0^+$ \cite{Camara:2022gvc}.
If we set $k=1$ and choose $b = \frac12 e^a$, we obtain the matrix that was
studied recently in \cite{Penna:2021kua} by a different method. 
We define the monodromy matrix ${\cal M}_{\rho,v} (\tau)$ by composition of \eqref{monpesuper} with 
\bea
\omega =  v + \frac{\rho}{2} \, \frac{1 +  \tau^2}{\tau}  \;\;\; (\lambda = -1) \;.
\eea
Note that 
each diagonal element of ${\cal M}_{\rho,v} (\tau)$ has essential singularities at $\tau =0$ and $\tau = \infty$. However, these
singularities do not constitute a problem in a WH factorisation approach, since the diagonal elements in ${\cal M}_{\rho,v} (\tau)$ are non-vanishing
H\"older continuous, indeed analytic, on any admissible
contour $\Gamma$ and admit a canonical WH factorisation given by \eqref{gfpm}-\eqref{gpm}.

{\remark \label{conhom}
 Note that all admissible contours
are homotopic in the punctured complex plane $\mathbb{C} \backslash \{0\}$, where ${\cal M}_{\rho,v}   (\tau)$ is analytic, and so we can take $\Gamma$, in this case,
to be any admissible contour.\\ 
}

Thus we obtain
\bea
{\cal M}_{\rho,v} (\tau) = {\rm diag} (e_-, e_-^{-1}) \, {\rm diag} (e_+ , e_+^{-1}) 
\eea
with
\bea
e_\pm=\exp \,P^\pm_{\Gamma} \left( f ( v + \frac{\rho}{2} \, \frac{1 +  \tau^2}{\tau} )\right) \;\;\;,\;\;\; f(\omega) = 4 b \, e^{- a k}  \cos  \left( k  \omega \right) , 
\eea
from which, upon normalisation of the factor $( M_{\rho,v})_+ (\tau)$ to the identity at $\tau = 0$, we get the solution
\bea
M_{ER} (\rho, v) = {\rm diag} \left( e^{f(v) J_0(k \rho)}, e^{ - f(v) J_0(k \rho)} \right) \;.
\label{MER}
\eea
Here we used the integral representation of the Bessel function of the first kind $J_0$,
 \bea
 J_0 (\rho) = \frac{1}{2 \pi i}  \int_{\Gamma} \,
 \frac{  e^{  \rho \, \left( z - \frac{1}{z}    \right)/2 } }{z} \, dz \;,
 \eea
 which satisfies $J_0 (- \rho) = J_0 (\rho)$.

 The associated four-dimensional solution \eqref{4dWLP} (with $\lambda = -1$) is
\bea
ds_4^2 =  \Delta \, dy^2 + \Delta^{-1}
\left(  e^\psi \, \left(  d \rho^2 -  dv^2 \right) + \rho^2 d\phi^2 \right) \;,
\label{ersol}
\eea
with
\bea
\Delta = e^{f(v) J_0(k \rho)} \;,
\eea
and
the metric factor $\psi (\rho, v)$ follows from \eqref{eq_psi} by integration,
\bea
\psi (\rho, v) = \left( 2 b \, e^{- a k} \right)^2  \left( k^2 \rho^2 J^2_0 (k \rho) + k^2 \rho^2 J^2_1 (k \rho) -
2 k \cos^2(k v)  \,   \rho \, J_0(k \rho) J_1 (k \rho) \right).
\eea
In the last equality
one integration constant has been dropped, and $J_1$ denotes a Bessel function of the first kind.

Setting $k=1$ and $b = \frac12 e^a$,
the metric \eqref{ersol} describes the Einstein-Rosen wave solution recently discussed in \cite{Penna:2021kua}.

As a consequence of Theorem \ref{theoMN}, taking also Remark \ref{rem:diag} into account, we have the following
remarkable property of the Einstein-Rosen solutions.

{\proposition \cite{Camara:2022gvc} \label{prop:mm}
Multiplying a solution of the form \eqref{MER} by any solution $M(\rho,v)$ of the field equations, 
with the same value $\lambda$ ($\lambda = -1$) and the
same line element $ds_2^2$, obtained from a pair $(M,X)$ satisfying the $\tau$-invariance property on an admissible contour, 
we obtain a new solution to the field equation \eqref{fi2d}.\\

} 

As an example thereof, we have the following.

{ \example  \underline{Deforming a Kasner solution:}

We return to the Kasner solution \eqref{Mkas} of Examples \ref{ex:cosms} and \ref{ex:kas}, where $\lambda = -1$, and interchange $\rho$ and $v$ in 
$ds_2^2 = dv^2 - d \rho^2 $ in \eqref{linrv},  so as to obtain $ds_2^2 = d \rho^2 - dv^2$ (which corresponds to taking $\sigma = 1, \varepsilon = -1$ in \eqref{2dl}, instead of 
 $\sigma = -1, \varepsilon = 1$ as in \eqref{linrv}).
 Note that in the Einstein-Rosen wave solution \eqref{ersol} we also have $ds_2^2 = d \rho^2 - dv^2$. 
Both the Kasner solution and the Einstein-Rosen wave solutions are represented by diagonal matrices, see Remark \ref{rem:diag}. Now recall that the Kasner solution
\eqref{Mkas}, which here we will denote by $M_K (\rho,v)$, satisfies the $\tau$-invariance property,
\bea
\frac{\partial}{\partial \tau} G_{M_K,X_K} (\tau, \rho, v) = 0 \;,
\eea
with $X_K = X$ given by \eqref{Xmer}, on a contour $\Gamma$ with $ \tau_0$, defined in \eqref{t0til}, in its exterior.
Applying Theorem \ref{theoMN} and Proposition \ref{prop:mm}, we thus obtain, taking $k=1$ and ${\tilde b} \equiv 2 b e^{-a}$ in \eqref{MER}, a new solution of the form 
\bea
ds_4^2 =  \Delta \, dy^2 + \Delta^{-1}
\left(  e^\psi \, \left(  d \rho^2 -  dv^2 \right) + \rho^2 d\phi^2 \right) 
\eea
given by
\bea
M(\rho,v) = M_{ER} (\rho,v) M_K (\rho) = {\rm diag} (\Delta, \Delta^{-1}) 
\eea
with
\bea
\Delta = \left( \frac{\rho}{2} \right)^4 \, e^{2 {\tilde b} \, \cos v \, J_0 (\rho)} \;.
\eea
The scalar function $\psi$ determined from \eqref{eq_psi} by integration is
\bea 
\psi (\rho, v) = - 2 \, \frac{J_0(\rho) \left( 2 - {\tilde b} \, J_1(\rho) \cos v \right)^2 }{\rho \, J_1(\rho)} + 
\int d\rho \, J_0^2(\rho) \left( - \frac{8}{\rho \, J_1^2 (\rho)} + 2 {\tilde b}^2 \rho \right) .
\eea
Using Proposition \ref{prop:mm}, this highly non-trivial solution is obtained  in a rather straightforward manner. It
can be viewed as arising from a deformation of a Kasner solution through multiplication by Einstein-Rosen solutions
(with deformation parameter ${\tilde b}$).

}

\section{Examples \label{sec:exam}}

We now present several concrete examples demonstrating that the WH factorisation method provides a practical means of obtaining explicit and exact solutions to the field equations of gravitational field theories.
Although no general methods exist for obtaining a canonical WH factorisation of a matrix function,
we show that by taking advantage of the particular structure of the matrix, simple tailored methods can be applied to determine the factorisation and to avoid the difficulties that arise, when other techniques are used, in the presence of multiple poles and essential 
singularities. In the examples below, we make use of the fact that the existence of a canonical WH factorization is stable under small perturbations (see Appendix \ref{sec:TRHWH}). Starting from a known solution and considering deformations of the associated monodromy matrix, we obtain new solutions. In particular, one finds that even small perturbations -- measured as small differences in the $L^{\infty}$-norm of the symbol (cf. Appendix  \ref{sec:TRHWH}) -- can lead to highly nontrivial changes in the space-time solution. 
We note that the study of hidden symmetries and integrable structures in the equations governing the dynamics of perturbed black hole solutions in four space-time dimensions is an active area of research, see \cite{Combaluzier-Szteinsznaider:2024sgb,Jaramillo:2024qjz} and references therein.

\subsection{Deforming the Schwarzschild monodromy matrix}

We begin by considering a deformation of the Schwarzschild metric, which arises from the canonical WH factorisation of the monodromy matrix 
 \cite{Cardoso:2017cgi} 
\begin{eqnarray}
{\cal M }_{\rho, v} (\tau) = 
\begin{pmatrix}
 \frac{\omega -m}{\omega +m}  \cosh \frac{\xi}{\omega}  & \sinh \frac{\xi}{\omega}  \\
\sinh \frac{\xi}{\omega}  & \frac{\omega +m}{\omega -m}  \cosh \frac{\xi}{\omega}  
\end{pmatrix}_{ \vert_{ \omega = v + \frac{ 1}{2}  \rho \frac{1 - \tau^2}{\tau} } }
\;\;\;,\;\;\; \det {\cal M } = 1 \;,
\label{Mmondef}
\end{eqnarray}
where $\xi \in \mathbb{R}$. When $\xi =  0$, it coincides with the Schwarzschild monodromy matrix \eqref{monschwarz} (with $\lambda = 1)$, but it ceases to be diagonal when $\xi \neq 0$.
Note that ${\cal M }_{\rho, v}$ is not a rational matrix. One can, however, take advantage of the fact that ${\cal M }_{\rho, v}$ is essentially a Daniele-Krapkov matrix \cite{ES}, taking the spectral parameter $\tau$
as the complex variable and $\rho,v$ as parameters. It can be decomposed as
\bea
{\cal M }_{\rho, v} = \Sigma D \Sigma^{-1} J \;,
\eea
with
\bea
\Sigma = \begin{pmatrix} 1 & 1\\ R & - R 
\end{pmatrix} \;\;\;,\;\;\; D = {\rm diag} \left( e^{\xi/\omega}, - e^{-\xi/\omega} \right)_{ \vert_{ \omega = v + \frac{ 1}{2}  \rho \frac{1 - \tau^2}{\tau} } } \;\;\;,\;\;\; 
J = \begin{pmatrix} 0 & 1 \\
1 & 0 
\end{pmatrix} \;,
\eea
where
\bea
R =  \left( \frac{\omega +m}{\omega -m}\right)_{ \vert_{ \omega = v + \frac{ 1}{2}  \rho \frac{1 - \tau^2}{\tau} } } \;.
\eea
We factorise ${\cal M }_{\rho, v} (\tau)$ with respect to the unit circle.
The scalar factorisation of $ \left(  e^{\pm \xi/\omega} \right)_{ \vert_{ \omega = v + \frac{ 1}{2}  \rho \frac{1 - \tau^2}{\tau} }} $ yields,
following \eqref{gfpm}-\eqref{gpm}, 
a canonical WH factorisation $D = D_- D_+$. 
Using this,  we obtain a RH problem for the unknown factors $(M_{\rho,v})_{\pm}$  in a canonical WH factorisation of ${\cal M }_{\rho, v} $, of the form
\bea
D_+ \Sigma^{-1} J (M_{\rho,v})_{+}^{-1} = D_-^{-1} \Sigma^{-1} (M_{\rho,v})_{-}\;.
\label{DrelD}
\eea
Since both sides of \eqref{DrelD} are meromorphic, one can apply a generalisation of Liouville's theorem to solve the RH problem in terms of an algebraic linear system, with coefficients
depending on $\rho,v$ as parameters, for two unknowns $K_1 (\rho,v), K_2 (\rho, v)$ (cf. \cite[Section 3.2.1]{Cardoso:2017cgi}). The resulting matrix $M(\rho,v)$ encoding the space-time solution, as in \eqref{MmMrv}, 
takes the form 
\begin{eqnarray}
M(\rho, v)  = \tfrac12 
\begin{pmatrix}
K_1 \qquad
& \qquad  K_2
\\
K_2  \qquad & \qquad  
 \frac{ 4 + K_2^2}{ K_1}
\end{pmatrix} \;.
\label{Mdf}
\end{eqnarray}
The associated space-time solution is described by a stationary line element of the form
\bea
ds_4^2 = -  \Delta (dt  + B  d \phi  )^2 + \Delta^{-1} 
\left(  e^\psi \, \left( d \rho^2 + dv^2 \right) + \rho^2 d\phi^2 \right) \;,
\eea
where
\bea
\Delta (\rho,v) = \frac{K_1 (\rho,v)}{4 + K_2^2 (\rho,v)} \;,
\eea
while $B$ and $\psi$ are determined using \eqref{M22} and \eqref{eq_psi}, respectively. To first order in the deformation parameter $\xi$ we have \cite{Cardoso:2017cgi}
\begin{eqnarray}
\Delta &=& \frac{\tau_2^+}{\tau_1^+} + {\cal O} (\xi^2) \;, \nonumber\\
B &=&2 \xi \, \frac{  \tau_2^+}{\rho( \tau_0^+ - \tau_0^-)   \, \tau_1^+
 (1 + \tau_1^+ \,  \tau_2^+) (\tau_1^+ - \tau_0^-) (\tau_2^+ - \tau_0^-) } \nonumber\\
&& \times \Big(  \tau_2^+ \,  \tau_0^-  - (\tau_0^-)^2 + \tau_1^+ (\tau_2^+ + \tau_0^-) 
(1 + \tau_2^+ \tau_0^-) -
 (\tau_2^+)^2 (1 + (\tau_0^-)^2)  \nonumber\\
&& \qquad  -  
   (\tau_1^+)^2 (1 + (\tau_2^+)^2  
      - (\tau_2^+) \, \tau_0^-   + (\tau_0^-)^2) \Big)  + {\cal O} (\xi^2)  \;,
   \end{eqnarray}
while $\psi$ is undeformed at first order in $\xi$. Here,
\bea
\tau_0^{\pm} &=& \frac{v \pm \sqrt{v^2 + \rho^2}}{\rho} \;, \nonumber\\
\tau_1^+ &=& \frac{(v-m) + \sqrt{(v-m)^2 + \rho^2}}{\rho} \;\;\;,\;\;\; \tau_2^+ = \frac{(v+m) + \sqrt{(v+m)^2 + \rho^2}}{\rho} \;.
\eea

\subsection{Deforming a static attractor solution \label{sec:nsol}}
%%%%%%%%%%%%%%%%%%%%%%

The field equations of the four-dimensional Einstein-Maxwell-dilaton theory, obtained via Kaluza-Klein reduction of five-dimensional Einstein gravity, admit extremal black hole solutions that exhibit the attractor mechanism
\cite{Astefanesei:2006dd}.
As an example, we consider a static extremal black hole. Its near-horizon solution is
given in terms of a four-dimensional $AdS_2 \times S^2$ space-time, whose line element in spherical coordinates reads
\bea
ds^2_4 = - \frac{r^2}{Q P } dt^2 + Q P  \left(\frac{dr^2}{r^2} + d \theta^2 + \sin^2 \theta \, d\phi^2 \right) \;.
\eea
The near-horizon solution is supported by the electric-magnetic field strength $F$  and by a constant scalar field $e^{- 2 \Phi} = Q/P$  (the dilaton field), where $Q$ and $P$ denote an electric and a magnetic charge, respectively.
We take $Q, P > 0$. The conversion to Weyl coordinates $(\rho,v)$ is made using  $\rho = r \sin \theta, v = r \cos \theta$.

As shown in \cite[Section 8.2.2]{Camara:2017hez}, upon reduction to two dimensions, this near-horizon solution is encoded in a matrix $M \in G/H$, where $G/H = SL(3, \mathbb{R})/SO(2,1)$. This matrix $M(\rho,v)$ 
results form the canonical WF factorisation (with respect to the unit circle) of a monodromy matrix  $ {\cal M }^{\rm seed}_{\rho, v} (\tau)  $, obtained from 
\bea
 {\cal M }^{\rm seed} (\omega)  = 
 \frac{1}{\omega^2} \begin{pmatrix}
A  & \quad B \omega   &\quad C  \omega^2
 \\
- B \omega  & D  \omega^2
 &  0
  \\
  C \omega^2 &
0
   & 0
\end{pmatrix} \;\;\;,\;\;\; \det {\cal M }^{\rm seed} = 1 \;,
\label{underrotmonvg0}
\eea
by substituting $\omega = v + \frac{\lambda}{2} \rho \frac{\lambda - \tau^2}{\tau}$ with $\lambda = 1$. 
The constants $A, B, C, D$  are expressed in terms of $Q, P$ by
\begin{eqnarray}
\label{valuesABCD}
A = - \frac{B^2}{2 D} \;\;\;,\;\;\; 
B= \frac{1}{2 \sqrt{\pi}} \, P^{1/3} \, Q^{2/3} \;\;,\;\; C = - \left(\frac{P}{Q} \right)^{1/3}
 \;\;,\;\; D = - \left(\frac{Q}{P} \right)^{2/3} \;.  
\end{eqnarray}
The resulting monodromy matrix $ {\cal M }^{\rm seed}_{\rho, v} (\tau)  $
can be deformed in different ways.
If the deformed monodromy matrix possesses a canonical WH factorisation, as is the case when the deformation parameters are small enough (cf. Appendix \ref{sec:TRHWH}), then this factorisation
will yield a solution to the field equations of the theory. In \cite[Section 10]{Camara:2017hez}, a deformation of the form
\bea
 {\cal M } (\omega)  = 
 \frac{1}{\omega^2} \begin{pmatrix}
A  & \quad B \omega + \alpha   &\quad C  \omega^2
 \\
- B \omega - \alpha  & D  \omega^2
 &  0
  \\
  C \omega^2 &
0
   & 0
\end{pmatrix} 
\label{aldef}
\eea
was considered, yielding a solution with interesting features. In particular it was observed that, although the deformation of the monodromy matrix \eqref{aldef}  is linear in $\alpha$, the corresponding space-time
solution receives corrections that are of order $\alpha^2$.

In the following, we focus on a different deformation \cite{Cardoso:2017cgi}, by replacing $Q$ and $P$ in \eqref{underrotmonvg0} by
\bea
Q \rightarrow Q + h_1 \omega \;\;\;,\;\;\; P \rightarrow P + h_2 \omega \;,
\eea
where we view $h_1, h_2  \in \mathbb{R}^+$ as deformation parameters. Thus, we obtain
\begin{eqnarray}
{\cal M}(\omega)  =
\left( \frac{H_2}{H_1} \right)^{1/3} 
  \begin{pmatrix}
H_1 H_2   &\quad  \sqrt{2} H_1  &\quad  - 1 \\
- \sqrt{2} H_1  &\quad  - H_1/H_2 
& \quad  0 \\
- 1 &\quad  0 & \quad 0 
\end{pmatrix} \;\;\;,\;\;\; \det {\cal M} = 1 \;,
\label{monomatrixMHH}
\end{eqnarray}
where
\bea
H_1 (\omega)  = h_1 + \frac{ Q}{\omega} \;\;\;,\;\;\;  H_2 (\omega) =  h_2 + \frac{ P}{\omega} \;.
\eea
There are several classes of contours that one may pick to perform a canonical WH factorisation of the matrix ${\cal M}_{\rho,v} (\tau)$, obtained from \eqref{monomatrixMHH}
by substituting $\omega = v + \frac{1}{2} \rho \frac{1 - \tau^2}{\tau}$, which may result in different solutions to the field equations, that were briefly discussed in \cite{Aniceto:2019rhg}.
We choose an admissible contour $\Gamma$ (with fixed points $\pm i$, since we take $\lambda = 1$), such that the following three poles (among a total of six) of ${\cal M}_{\rho,v} (\tau)$, 
\begin{eqnarray}
\tau_{0  }&=& \frac{1}{\rho}\Big (v -  \sqrt{\rho^2 + v^2}\Big ) \;, \nonumber\\
\tau_{\tilde Q } &=& \frac{1}{\rho} \left(v + {\tilde Q}
- \sqrt{\rho^2 + (v+ {\tilde Q})^2} \right) \;\;\;,\;\;\; {\tilde Q} = \frac{Q}{h_1}\;,  \nonumber\\
\tau_{\tilde P } &=& \frac{1}{\rho} \left(v + {\tilde P} - \sqrt{\rho^2 + (v+{\tilde P})^2} \right)  \;\;\;,\;\;\; {\tilde P} = \frac{P}{h_2} \;,
\label{tval}
\end{eqnarray}
lie inside $\Gamma$, as in \cite{Cardoso:2017cgi,Aniceto:2019rhg}. Note that these three poles are real and negative. By Corollary \ref{cor:iii} (ii), the monodromy matrix ${\cal M}_{\rho,v} (\tau)$ admits a canonical WH factorisation.
In this case, ${\cal M}_{\rho,v} (\tau)$ is a rational matrix, so a canonical WH factorisation can be obtained following the systematic procedure described in \cite[Section 3]{Camara:2017hez}, by using a generalisation of Liouville's theorem.
Note that, by using this method, the appearance of double poles does not constitute a problem, nor does it lead to significant computational difficulties, unlike in the case of other RH approaches (see for example \cite[Section 3.1]{Katsimpouri:2013wka}).

We now summarise the results of \cite{Aniceto:2019rhg} for this case.
When 
${\tilde Q} = {\tilde P}$, the solution resulting from the canonical factorizaton of ${\cal M}_{\rho,v} (\tau)$ describes a four-dimensional
extremal black hole supported by a constant dilaton field. We therefore take ${\tilde Q} \neq {\tilde P}$, with  ${\tilde P} > {\tilde Q} >0$ for definiteness.
The resulting expressions 
for  $\Delta, B, e^{\psi}$ in the line element
\begin{equation}
ds^2_4 = - \Delta 
\, \left(dt + B \, d\phi \right)^2 + \Delta^{-1}  \left(e^{\psi} \, (d \rho^2 + dv^2) + 
\rho^2 \, d\phi^2 \right) \;,
\label{linePQg0}
\end{equation}
for the electric-magnetic field strength $F$ and for the dilaton field $e^{- 2 \Phi}$ are given in \cite[Appendix A]{Aniceto:2019rhg}.
The metric \eqref{linePQg0} has two Killing horizons, $ || \partial/ \partial t||^2 = \Delta = 0$,
located at $\rho = v = 0$ and at $\rho = 0, - {\tilde P} < v < - {\tilde Q}$, respectively.
When approaching the 
Killing horizon $\rho = v = 0$, keeping  $\rho/v$ constant,
the metric takes the form
\begin{equation}
ds^2_4 = - \frac{\rho^2 + v^2}{P Q} 
\, \left(dt +  h_1 h_2 \, {\tilde J} \, f\left(v/\sqrt{\rho^2 + v^2} \right) \,  d\phi \right)^2 + \frac{PQ}{\rho^2 + v^2}  \left( d \rho^2 + dv^2 + 
\rho^2 \, d\phi^2 \right) \;,
\end{equation}
where $f(x) = x (1-x) -1$ denotes a linear combination of the Legendre polynomials $P_0, P_1$ and $P_2$. 
The scalar field approaches the value $e^{- 2 \Phi} \rightarrow P/Q$. Note, however, that $\partial_{\rho,v} e^{- 2 \Phi} $
does not vanish at $\rho=v=0$ when $\tilde J \neq 0$. Thus, only when ${\tilde J} =0$ 
(that is, in the extremal black hole case)
does the solution exhibit an attractor behaviour as
one approaches $\rho=v=0$. 
Both the Ricci and the Kretschmann scalars are well-behaved at the Killing horizon $\rho=v=0$.
However, they both blow up at the Killing horizon $\rho = 0, \, -\tilde{P}< v < -\tilde{Q}$, which points to the existence of a curvature singularity at this horizon.
The scalar field 
$e^{- 2 \Phi}$ also diverges at this Killing horizon.

As $\rho^2 + v^2 \rightarrow + \infty$, the solution asymptotes to a stationary solution with an effective NUT parameter $\tilde J = {\tilde P} - {\tilde Q}$
which is expressed in terms of the electric-magnetic charges,
\bea
ds^2_4 &=& - \frac{1}{h_1 h_2} \left(1 - \frac{{\tilde P} + {\tilde Q}}
{\sqrt{\rho^2 + v^2}} \right)
\, \left(dt - h_1 h_2 \, {\tilde J} \, \frac{v}{\sqrt{\rho^2 + v^2}} \, d\phi \right)^2 \nonumber\\
&& + h_1 h_2  \left(1 + \frac{{\tilde P} + {\tilde Q}}
{\sqrt{\rho^2 + v^2}} \right)
\left(d \rho^2 + dv^2 + 
\rho^2 \, d\phi^2 \right) \;,
\eea
with $e^{- 2 \Phi} \rightarrow h_1/h_2$.

Thus, this solution describes a space-time that is supported by one electric and one magnetic charge and by a scalar field, possesses two Killing horizons and asymptotes
to a space-time with an effective NUT parameter $\tilde J = {\tilde P} - {\tilde Q}$ that is expressed in terms of the electric-magnetic charges.
 This solution has similarities
with a solution discovered by Brill \cite{Brill} (see also \cite{Stephani:2003tm}) in a different four-dimensional theory, namely an Einstein+Maxwell theory, in
that it possesses two Killing horizons, one of them being associated with the presence of a NUT parameter. However, 
while Brill's solution describes an electrically charged (or magnetically charged) Reissner-Nordstrom black hole when the NUT parameter is switched off, the solution discussed above
describes a dyonic extremal black hole solution (that is supported by a scalar field) when the NUT parameter $\tilde J$ is set to zero. Moreover,
differently from Brill's solution, the NUT parameter
${\tilde J}$ is not an additional parameter, but rather an effective parameter that is expressed in terms of the electric-magnetic charges.

\section{Open questions \label{sec:open}}

To conclude, we list a few open questions related to the subjects covered in this review.

\begin{enumerate}

\item Which monodromy matrices should one pick to generate solutions to the gravitational field equations with specific physical properties \cite{Roy:2018ptt}?
For instance, how do choices of rod structures \cite{Harmark:2004rm,Emparan:2001wk,Katsimpouri:2014ara}  get reflected in the choice of monodromy matrices \cite{Roy:2018ptt}?
And how do hidden integrable structures in the equations governing the dynamics of perturbed black hole solutions \cite{Combaluzier-Szteinsznaider:2024sgb} get reflected in the choice of monodromy matrices?

\item Which classes of pairs of matrix functions $(M,X)$ satisfy the property of $\tau$-invariance \eqref{tderGt}? Do all the solutions of \eqref{fi2d} and the associated linear system \eqref{laxx} satisfy the property of $\tau$-invariance \eqref{tderGt}?

\item Are there other types of matrix factorisations which yield solutions to  \eqref{fi2d}?
 
\item What is the relation of the linear system \eqref{laxx}  with the linear system description of self-dual Yang-Mills theories, as suggested by the classical double copy
(which relates solutions of the Yang–Mills equations to solutions of Einstein’s equations) (cf. \cite{LopesCardoso:2024ttc})?

\end{enumerate}

%%%%%%%%%%%%%%%%%%%

\section*{Acknowledgements}
Research partially funded by Funda\c{c}\~ao para a Ci\^encia e Tecnologia (FCT), Portugal, through grant No. UID/4459/2025.}

\appendix

\section{Toeplitz operators, RH problems and WH factorisation \label{sec:TRHWH}}

General references for this section are \cite{CG,MP,CDR,LS,GK,CCT}.

Let $\gamma$ be a simple closed path around the origin, let $\mathbb{D}^+_{\gamma} = {\rm int} \gamma$, 
$\mathbb{D}^- _{\gamma} = {\rm ext} \gamma$ and let $L^2 (\gamma)$ denote the space of all square-integrable functions on $\gamma$.  We denote by 
$S_{\gamma}$ the singular integral with Cauchy kernel 
\bea
  \left( S_{\gamma} \varphi \right) (\tau) = \frac{1}{ \pi i} \mathrm{P.V.}  \int_{\gamma} \, \frac{\varphi(u)}{u - \tau} \, du  \;\;\;,\;\;\; \tau \in \gamma \;,
 \eea
 where  $\mathrm{P.V.}$ denotes Cauchy's principal value. $S_{\gamma}$ 
defines a bounded operator on $L^2 (\gamma)$ and satisfies
\bea
S^2_{\gamma} = \textrm{Id} \;,
\eea
where $\textrm{Id}$ denotes the identity operator.
This allows one to define two complementary projections,
\bea
P^{\pm}_{\gamma} = \frac12 \left( \textrm{Id} \pm S_{\gamma} \right) \;,
\eea
inducing an orthogonal decomposition of $L^2 (\gamma)$,
\bea
L^2 (\gamma) = H^2_+ \oplus H^2_- \;\;\;,\;\;\; H^2_{\pm} : = P_{\gamma}^{\pm} L^2 (\gamma) \;.
\eea
The subspaces $H^2_{\pm}$ can be identified with spaces of analytic functions in $\mathbb{D}^{\pm}_{\gamma} $, respectively, as being their non-tangential boundary value functions, defined a.e. on $\gamma$ \cite{Duren,Hoffman}. 
For any $f_- \in H^2_- $ we have that $f_- (\infty) = 0$.

Now let $h$ be an essentially bounded function on $\gamma$ and define
\bea
T_h : H^2_+ \rightarrow H^2_+ \;\;\;,\;\;\;\; T_h f_+ = P_{\gamma}^+ \left( h f_+ \right)
\;\;\;,\;\;\; f_+ \in H^2_+ \;. 
\label{Tg}
\eea
$T_h$ is a bounded operator on $H^2_+$ and is called a {\it Toeplitz operator}; $h$ is called the  {\it symbol} of the operator.

Matricial (also called block) Toeplitz operators can be defined analogously in $\left( H^2_+ \right)^n$ if $h$ is an 
$n \times n$ essentially bounded matrix function. In that case,  $P^+$ is applied componentwise in \eqref{Tg}, with
$ f_+ \in \left( H^2_+ \right)^n$.
The operator norm of $T_h$ is proportional to the $L^{\infty}$-norm of its symbol, 
\bea
||T_h|| = c \,   ||h||_{\infty} \;,
\label{Thnorm}
\eea
where, for $h=[h_{ij}]_{i, j= 1, \dots, n}$,
\bea
 ||h||_{\infty} = n \max_{i,j} ||h_{ij}||_{\infty} \;\;\;,\;\;\;  ||h_{ij}||_{\infty} = {\rm ess} \sup_{t \in \gamma}  |h_{ij} (t)| \;.
 \eea

The kernels of Toeplitz operators have attracted considerable attention owing to their significance as spaces of analytic functions and their distinctive properties 
 \cite{sarrason,hartmann2015,Nakazi,GarciaMashreghiRoss2016}.
Noting that $P_{\gamma} ^- = \textrm{Id} - P_{\gamma}^+$, it follows from the definition \eqref{Tg} that
\bea
\ker T_h &=& \{ f_+ \in \left( H^2_+ \right)^n: T_h f_+ = 0 \} = 
\{ f_+ \in \left( H^2_+ \right)^n: P_{\gamma}^+ \left( h f_+  \right) = 0 \} \nonumber\\
&=& 
\{ f_+ \in \left( H^2_+ \right)^n: h f_+ \in  \left( H^2_- \right)^n \} \;.
\eea
So, the kernel of a Toeplitz operator $T_h$ consists of the solutions $f_+ \in \left( H^2_+ \right)^n$ of the (vectorial) RH problem with matricial coefficient $h$, of the form
\bea 
h f_+ = f_- \;\;\;,\;\;\; f_{\pm} \in \left( H^2_+ \right)^n \;.
\label{gfpfm}
\eea
We can thus express the injectivity of $T_h$ in terms of a vectorial RH problem.

{\proposition \label{prop·inj} 
$T_h$ is injective if and only if \eqref{gfpfm} admits only the zero solution. 
\\
}

Fredholmness of Toeplitz operators is also closely related with matricial RH problems.
Denoting the adjoint of an operator $T$ on a Hilbert space $\cal H$ by $T^*$, we say that $T$ is Fredholm 
 if and only if $\dim \ker T < \infty$,
 $\dim \ker T^* < \infty$ and $T$ has a closed range ${\rm Im} T$. Note that
 $\dim \ker T^* = {\rm codim} \, {\rm Im} T : =  { \dim {\cal H}}/ \overline{{\rm Im} T }$ \cite{MP, GK}. 
 A necessary and sufficient condition for a Toeplitz operator to be Fredholm is the following.

 {\theorem
 A Toeplitz operator $T_h$ on $\left( H^2_+ \right)^n$  is Fredholm if and only if $h$ admits a WH factorisation
 \bea
 h = h_- \, {\rm diag} \left( \tau^{k_j} \right)_{j = 1, \dots, n} \, h_+  \quad \text{on} \quad \gamma \;,
 \label{gmptau}
 \eea
 where 
 \bea
  h_-^{\pm 1} \in \left( H_-^2 \oplus \mathbb{C} \right)^{n \times n} \;\;\;,\;\;\; 
  h_+^{\pm 1} \in \left( H_+^2  \right)^{n \times n} 
  \label{hphm}
  \eea
  and  $k_j \in \mathbb{Z}$ for all $ j = 1, \dots, n$.\\
 \\
 }

Note that if a representation of the form \eqref{gmptau} exists, the exponents 
 $ k_j $ (called partial indices of $h$) 
 are uniquely defined up to their order and 
 \bea
 \dim \ker T_h = \sum_{k_j < 0} |k_j| \;\;\;,\;\;\;  {\rm codim} \, {\rm Im} T_h  = \sum_{k_j > 0} k_j \;.
 \eea
 The sum  of all indices $k_j$ is called the {\it total index of $h$}, denoted ${\rm ind}_{\gamma} (h) $, and we have that 
\bea
\dim \ker T_h -  {\rm codim} \, {\rm Im} T_h = - {\rm ind}_{\gamma} (h) \;.
\label{dimkTcT}
\eea

If all the elements of the matrix symbol $h$ belong to the algebra of H\"older continuous functions on $\gamma$ with exponent $\mu \in \, ]0, 1[$, denoted
$C^{\mu}_{\gamma}$ \cite{MP},  and $\det h$ does not vanish on $\gamma$, then a WH factorisation \eqref{gmptau} always exists, with $h_-^{\pm 1} , h_+^{\pm 1} \in \left(C_{\gamma}^{\mu}\right)^{n\times n}$.
In that case, $ {\rm ind}_{\gamma} (h) $ coincides with the winding number of $\det h$ around the origin.
If $\det h = 1$, then $ {\rm ind}_{\gamma} (h) = 0$. We thus have, from \eqref{dimkTcT}: 

{\proposition \label{prop:Tinv}
A Toeplitz operator $T_h$ with $h$ in $\left(C_{\gamma}^{\mu}\right)^{n\times n} $ and $\det h = 1$ is Fredholm and 
\bea
 \dim \ker T_h =  {\rm codim} \, {\rm Im} T_h \;.
 \eea

}

\vskip 3mm
Since a Fredholm operator with $ \dim \ker T_h =  {\rm codim} \, {\rm Im} T_h =0$ is invertible, we conclude the following.

{\corollary \label{cortii}
With the same assumptions as in Proposition \ref{prop:Tinv}, 
\bea
T_h \; \; \text{is \; injective} \implies T_h \;\; \text{is \; invertible} 
\label{cor:Tii}
\eea

}
\vskip 3mm

Since the converse of \eqref{cor:Tii} is clearly true, we see that for matrix symbols satisfying the conditions of Proposition \ref{prop:Tinv}, invertiblity is equivalent to injectivity, i.e., to \eqref{gfpfm}
admitting only the trivial solution.

Invertibility of a Toeplitz operator can also be expressed in terms of WH factorisation, as follows.

{ \theorem  \label{theo:Tinvc}
The operator $T_h$ is invertible in $\left( H_+^2 \right)^n$ if and only if $h$ admits a canonical WH factorisation, i.e.
 \eqref{gmptau}-\eqref{hphm} hold with $k_j =0$ for all $j=1, \dots, n$. \\
 }

 As a consequence of Theorem \ref{theo:Tinvc}, Corollary \ref{cortii} and Proposition \ref{prop·inj}
 we thus have:

{\corollary

Let $h \in (C_{\gamma}^{\mu})^{n \times n}$ with $\det h = 1$. Then $h$ admits a canonical WH factorisation with respect to $\gamma$ if and only if
\bea
h f_+ = f_- \;\;\;,\;\;\; f_{\pm} \in ( H_{\pm}^2)^n \;,
\eea
admits only the trivial solution $f_{\pm} = 0$.\\
}

Since the set of all invertible operators is an open set in the space of all bounded linear operators on a Banach space, the invertibility property is stable under small
perturbations. {From} Theorem \ref{theo:Tinvc} and \eqref{Thnorm}, it follows that a similar property holds regarding the existence of a canonical WH factorisation. We formulate
it as follows.

{\corollary 
The existence of a canonical WH factorisation for a matrix function is stable under small perturbations of the matrix in the $L^{\infty}$-norm.

}

%%%%%%%%%%%%%%%%%%%%%%%%%%

\section{Canonical WH factorisation and Ward's factorisation approach \label{sec:wardcan}}

Consider the  four-dimensional line element \eqref{4dWLP}
with $\lambda = 1$,
\bea
ds_4^2 = - \Delta (dt  + B  d \phi  )^2 + \Delta^{-1} 
\left(  e^\psi \, \left( d\rho^2 + dv^2 \right)  + \rho^2 d\phi^2 \right) \;\;\;,\;\;\; \rho > 0 \;,
\label{4dWLP2}
\eea
where $\Delta > 0$. In the case of Einstein's field equations in vacuum in four space-time dimensions, 
the associated $2 \times 2$ matrix $M(\rho,v)$ encoding the line element \eqref{4dWLP2} takes the form
\bea
\label{M222}
M =
\begin{pmatrix}
 \Delta + {\tilde B}^2/\Delta &\;  {\tilde B}/ \Delta\\
{\tilde B}/\Delta  &\;  1/\Delta
\end{pmatrix} \;,
\eea
with
\bea
 \rho \star d {\tilde B} = \Delta^2 \, dB \;.
 \label{BBt}
\eea
Defining 
\bea
e^{\Gamma} = \frac{e^{\psi}}{\Delta} \;,
\eea
the line element \eqref{4dWLP2} can be written as 
\bea
ds_4^2 = - \Delta dt^2 - 2 \Delta B \, dt \, d \phi + \left( \Delta^{-1} \rho^2 - \Delta \, B^2 \right) d \phi^2 + e^{\Gamma} 
 \left( d\rho^2 + dv^2 \right) \;,
\eea
where the first three terms on the right-hand side can be represented by a $2 \times 2$ symmetric matrix $g$, with $\det g = -1$, 
\bea
g = - \frac{1}{\rho} \begin{pmatrix}
\Delta \; & \Delta  B \\
\Delta B \; & \Delta B^2 - \frac{\rho^2}{\Delta}
\end{pmatrix} \;.
\label{matg}
\eea
Thus, 
the line element \eqref{4dWLP2} can also be written as \cite{Ward1982,Korotkin:2023lrg}
\bea
ds^2 = \rho \, g_{ij} dx^i dx^j + e^{\Gamma} \left( d \rho^2 + dv^2 \right) \;\;\;,\;\;\; i, j = 1,2 \;,
\eea
with $dx^1 = dt, dx^2 = d \phi$, and represented by the matrix $g(\rho,v)$.

Note that the two matrices $g$ and $M$ are defined in terms of different functions. Namely, while $g$ is defined in terms of the functions $\Delta$ and $B$,
$M$ is defined in terms of $\Delta$ and $\tilde B$, where $\tilde B$ is related to $B$ through the relation \eqref{BBt}.

Both $g$ and $M$ are obtained from a canonical WH factorisation of matrices depending on the complex parameter $\tau$ and the space-time coordinates $(\rho,v)$. The
matrix $g$ in Ward's construction \cite{Ward1982} is obtained from a canonical WH factorisation, with respect to $\tau$ on the unit circle $|\tau| = 1$, of a (non-symmetric) matrix
 \bea
H(\tau, \rho,v) = \begin{pmatrix}
    h_1 (\omega) & (- \tau)^k \, h_2 (\omega) \\
    \tau^{-k}  \, h_2 (\omega) & h_3 (\omega)
\end{pmatrix} \quad \text{with} \quad \omega= v + \frac{\rho}{2}  \left( \frac{ 1 - \tau^2}{\tau} \right) \;\;\;,\;\;\; \det H = -1 \;.
\label{Hwt}
\eea
Denoting the factors in the factorisation \eqref{splitF} by  $\left( H_{\rho,v} \right)_- = {\hat H}, \left(H_{\rho,v} \right)_+ = H^{-1} $, we have
\bea
g = P \left( H_{\rho,v} \right)_-  (\infty) \left(H_{\rho,v} \right)_+ (0) \, P \;,
\eea
where 
\bea
P = {\rm diag} \left( \rho^{-k/2} , \rho^{k/2} \right)  \;\;\;,\;\;\; k \in \mathbb{Z} \;.
\eea
In its turn, the matrix $M(\rho,v)$ in \eqref{M222} is obtained from the canonical WH factorisation of a (symmetric) matrix ${\cal M} (\omega)$, as described 
in Theorem \ref{theoraccr}.

The relation between Ward's construction \cite{Ward1982} and the one presented in Theorem \ref{theoraccr} is not immediate, since they are based on very different approaches,
but they can be shown to coincide when $h_2 =0$ in \eqref{Hwt}, which corresponds to setting $B = {\tilde B} =0$, in which case  $H(\tau, \rho,v)$ is a diagonal matrix that only depends on $\omega$.
Introducing the diagonal matrix $D = {\rm diag} \left( -1 , 1 \right) $, we define
\bea
{\tilde M} = D \, M =    {\rm diag} \left( - \Delta , \frac{1}{\Delta}  \right) 
 \;\;\;,\;\;\;  \det {\tilde M}  = -1\;.
\eea
We then have, for $k=1$, 
\bea
g = P^{-1} {\tilde M} P^{-1}  =  {\rm diag} \left( - \frac{\Delta}{ \rho} , \frac{\rho}{\Delta}  \right) \;.
\label{gDel}
\eea
If $M$ is obtained by canonical WH factorisation, with respect to the unit circle, of a monodromy matrix ${\cal M}_{\rho,v} (\tau)$, we have, following  \cite{Ward1982}, 
\bea
 {\cal M}_{\rho,v} (\tau) = D^{-1} \, {\hat H} (\tau) \, H^{-1} (\tau) =  \left( D^{-1} \, {\hat H} (\tau) \, H^{-1} (0) \right) \left( H(0)  \, H^{-1} (\tau) \right) \;,
 \eea
and, by Theorem  \ref{theoraccr} and \eqref{gDel},
\bea
M(\rho,v) &=&   D^{-1} \, {\hat H} (\infty) \, H^{-1} (0) \;, \nonumber\\
{\tilde M} (\rho,v) &=&  {\hat H} (\infty) \, H^{-1} (0) \;, \nonumber\\
g &=& P^{-1} \, {\hat H} (\infty) \, H^{-1} (0) \, P^{-1} \;,
\eea
which is the solution in signature $(-,+,+,+)$ that corresponds to the one given in 
\cite{Ward1982}.

%%%%%%%%%%%%%

\providecommand{\href}[2]{#2}\begingroup\raggedright\endgroup

%%%%%%%%%%%%%%%%%%%%%%%%%%

\end{document}